\documentclass[11pt,a4paper,english]{article}
\usepackage{graphicx}
\usepackage{amssymb,latexsym}
\usepackage{amsmath}
\usepackage{epsf}
\usepackage{pdfsync}
\usepackage{epsfig}
\usepackage[parfill]{parskip}
\usepackage[matrix,arrow,color]{xy}
\usepackage[usenames]{color}
\usepackage{hyperref}
\usepackage{mathrsfs}
\usepackage[mathcal]{euscript}
\usepackage[font={small}]{caption}
\usepackage{subcaption}
\usepackage[export]{adjustbox}
\usepackage{float}
\usepackage{wrapfig}
\usepackage{microtype}
\usepackage{slashed}
\usepackage [latin1]{inputenc}

\input{xy}
\xyoption{all}

\input{epsf}

\makeatletter

\usepackage{setspace}

\usepackage{lscape}

\catcode`@=11
\renewcommand{\section}
{\@startsection{section}{1}{0pt}{\medskipamount}{\medskipamount}{\large\bf}}
\makeatletter\renewcommand{\subsection}
{\@startsection{subsection}{2}{\z@}{-3.25ex plus -1ex minus -.2ex}
{1.5ex plus .2ex}{\it }}

\numberwithin{equation}{section}
\catcode`@=12

\newcommand{\ban}{\begin{eqnarray}}
\newcommand{\ean}{\end{eqnarray}}

\newcommand{\Tr}{{\rm Tr}}

\newcommand{\cI}{{\cal I}}
\newcommand{\cW}{{\cal W}}
\newcommand{\cN}{{\cal N}}
\newcommand{\cM}{{\cal M}}
\newcommand{\cS}{{\cal S}}
\newcommand{\cB}{{\cal B}}

\newcommand{\cH}{{\cal H}}

\newcommand{\cO}{{\cal O}}
\newcommand{\cC}{{\cal C}}

\newcommand{\cT}{{\cal T}}
\newcommand{\cK}{{\cal K}}

\newcommand{\sfQ}{{\mathsf{Q}}}

\newcommand{\sfX}{{\mathsf{X}}}

\newcommand{\sfW}{{\mathsf{W}}}

\newcommand{\scrF}{{\mathscr{F}}}

\newcommand{\Hecke}{\mathcal{H\kern-.77em H}}

\newcommand{\mbf}[1]{{\boldsymbol {#1} }}
\newcommand{\complex}{{\mathbb C}} 
\newcommand{\zed}{{\mathbb Z}} 
\newcommand{\real}{{\mathbb R}} 
\newcommand{\torus}{{\mathbb T}}

\def\dd{{\rm d}}

\newcommand{\Hom}{\mathrm{Hom}}

\newcommand{\motive}{\mathbb{L}}

\def\beq{\begin{equation}}
\def\bee{\begin{equation}}
\def\eeq{\end{equation}}
\def\bea{\begin{eqnarray}}
\def\eea{\end{eqnarray}}
\def\bd{\begin{displaymath}}
\def\ed{\end{displaymath}}

\newcommand{\Cint}{\int\kern-10.5pt-\kern7pt}

\newcommand{\PP}{{\mathbb{P}}}

\newcommand{\be}{\begin{equation}}
\newcommand{\ee}{\end{equation}}
\newcommand{\bal}{\begin{align}}
\newcommand{\eal}{\end{align}}

\newcommand\fverbit{\egroup\item[\fbox{\unhbox\pippobox}]}
\newbox\pippobox

{

\def\be{\begin{equation}}
\def\ee{\end{equation}}
\def\bea{\begin{eqnarray}}
\def\eea{\end{eqnarray}}

\begin{document}

\begin{titlepage}
\setcounter{page}{1}

\vskip 5cm

\begin{center}

\vspace*{3cm}

{\Huge Quantum Line Defects and \\[8pt] Refined BPS Spectra}

\vspace{15mm}

{\large\bf Michele Cirafici}
\\[6mm]
\noindent{\em Dipartimento di Matematica e Geoscienze, Universit\`a di Trieste, \\ Via A. Valerio 12/1, I-34127 Trieste, Italy, 
\\ Institute for Geometry and Physics (IGAP), via Beirut 2/1, 34151, Trieste, Italy
\\ INFN, Sezione di Trieste, Trieste, Italy 
}\\[4pt] Email: \ {\tt michelecirafici@gmail.com}

\vspace{15mm}

\begin{abstract}
\noindent

In this note we study refined BPS invariants associated with certain quantum line defects in quantum field theories of class $\cS$. Such defects can be specified via geometric engineering in the UV by assigning a path on a certain curve. In the IR they are described by framed BPS quivers. We study the associated BPS spectral problem, including the spin content. The relevant BPS invariants arise from the K-theoretic enumerative geometry of the moduli spaces of quiver representations, adapting a construction by Nekrasov and Okounkov. In particular refined framed BPS states are described via Euler characteristics of certain complexes of sheaves.

\end{abstract}

\vspace{15mm}

\today

\end{center}
\end{titlepage}


\tableofcontents


\section{Introduction}

Line defects, and extended objects in general, are part of the nonperturbative definition of quantum field theory. In this note we will consider line defects in supersymmetric theories from the perspective of enumerative geometry. Enumerative geometry provides a conceptual tool to understand the structure of the underlying moduli spaces. In physics language such moduli spaces are models for the Hilbert spaces of states of supersymmetric theories. Natural counting problems associated with their intersection theory correspond to Witten indices or generalizations thereof. 

In this note we will adapt the K-theoretic approach to the counting of BPS states developed by Nekrasov and Okounkov in \cite{Nekrasov:2014nea,Okounkov:2015spn} to study quantum line defects in class $\cS$ supersymmetric quantum field theories. The formalism of framed BPS quivers associates to a line operator a certain enumerative problem of Donaldson-Thomas type, the counting of certain stable quiver representations \cite{Chuang:2013wt,Cirafici:2018jor,Cirafici:2017wlw,Cordova:2013bza}. In certain cases this problem can be solved combinatorially via virtual localization with respect to a natural toric action. We will show that such a problem can be refined to include the spin information of the BPS states. The main technical tool we will use is the K-theoretic version of the localization computation as introduced in \cite{Nekrasov:2014nea,Okounkov:2015spn}. Under certain conditions, the K-theoretic construction produces certain toric modules out of which we can compute ordinary framed BPS states by taking Euler characteristics. This associates to the line defect a collection of rational functions (characters of $\torus$-modules). In certain cases, eventually by taking a scaling limit, such functions correspond to refined BPS degeneracies. The starting point of our construction are the results of \cite{Cirafici:2017wlw} which establish a formalism to compute framed BPS spectra associated with line defects using equivariant localization. This note improves on this result by showing how one can extend the localization computation to include spin/refined information.

Consider a path $\wp$ on a complex curve $\cC$, eventually endowed with additional data. String theory associates to $\wp$ a UV line operator in a four dimensional quantum field theory with $\cN=2$ on $S^1_R \times \real^3$, for example via the low energy description of a system of M5-M2 branes. In the IR such a line operator acquires an expectation value, which is a function on the moduli space of vacua of the theory \cite{Gaiotto:2010be,Moore:2015szp}
\be \label{introvev}
\langle \, L_\wp \, \rangle = \sum_{\gamma \in \Gamma} \ \underline{\overline{\Omega}} (\gamma ; \wp) \ X_\gamma
\ee
where $\underline{\overline{\Omega}} (\gamma ; \wp)$ are the degeneracies of framed BPS states, $\Gamma$ the lattice of electric and magnetic charges, and $X_\gamma$ are a collection of Darboux coordinates on the Hitchin moduli space of $\cC$ \cite{Gaiotto:2008cd}. In this note we give a prescription to replace this generating function for certain classes of line operators with the K-theoretic generating function
\be \label{introEC}
\mathscr{L}_\wp = \chi \left( \mathscr{M} , \cO^{\rm vir}_{\mathscr{M}} \otimes \cK^{1/2}_{\rm vir} \right)
\equiv
 \sum_{\gamma} \, \chi \left( \mathscr{M}_\gamma , \cO_{\mathscr{M}_\gamma}^{\rm vir} \otimes \mathcal{K}^{1/2}_{\rm vir} \right) \, \mathsf{X}_\gamma 
\ee
where $\mathscr{M} = \sqcup_\gamma \, \mathscr{M}_\gamma$ is a disjoint union of moduli spaces of quiver representations and $\cO^{\rm vir}_{\mathscr{M}} \otimes \cK^{1/2}_{\rm vir}$ is a certain sheaf. Hidden in \eqref{introEC} are certain formal variables $\sfX_\gamma$ which are placeholders for quantum coordinates on the geometric quantization of the Hitchin moduli space. This Euler characteristic captures the spin content of the framed BPS states in terms of certain refined Donaldson-Thomas type invariants. By replacing K-theory classes with ordinary Euler characteristics and the formal variables with actual coordinates $X_\gamma$, $\mathscr{L}_\wp$ reproduces \eqref{introvev}.

A companion paper will discuss the analog K-theoretic Donaldson-Thomas theory for quivers associated with Calabi-Yau singularities \cite{MQ}.

We will discuss the general background in Section \ref{generalities}. The main construction is presented in Section \ref{Ktheory}, while Section \ref{computations} contains several examples. We conclude with a short discussion in Section \ref{discussion}.

\section{Generalities} \label{generalities}

In this Section we provide some background to our construction. We discuss four dimensional quantum field theories of class $\cS$, their line defects and the associated framed BPS spectra.

\subsection{Line defects in theories of class $\cS$}

\paragraph{Theories of class $\cS$.} The name \textit{class $\cS$} refers to a class of theories which can be engineered via a compactification of the $\cN={(2,0)}$ superconformal theory in six dimensions on a curve $\cC$, down to four dimensions. The UV curve $\cC$ can have punctures and some extra data at the punctures. Such theories have moduli spaces of vacua; we will denote by $\cB$ their Coulomb branch, which parametrizes tuples of meromorphic differentials on $\cC$. At a generic point $u \in \cB$ the gauge symmetry is spontaneously broken down to the rank $r$ maximal torus of the gauge group $G$.

The low energy effective action for the theory is captured by the Seiberg-Witten curve $\Sigma_u$, a branched covering of $\cC$ \cite{Seiberg:1994rs}. Such a curve determines the lattice of charges $\Gamma = H_1 (\Sigma_u , \zed)$, endowed with an antisymmetric integral pairing $\langle \  , \ \rangle : \Gamma \times \Gamma \longrightarrow \zed$. The Seiberg-Witten differential $\lambda_u$ on $\Sigma_u$ determines the central charge function $Z_\gamma (u) = \frac{1}{\pi} \int_\gamma \lambda_u$ of a BPS state with charge $\gamma \in \Gamma$.

The central charge function poses a non-trivial BPS spectral problem over the Coulomb branch. The BPS Hilbert spaces decompose into superselection sectors $\cH_{\gamma, u}$ labelled by the charges $\gamma \in \Gamma$ and the spectral problem consists in determining the BPS degeneracies $\Omega (\gamma , u)$, certain indices computed over $\cH_{\gamma , u}$, as a function of $u \in \cB$.

\paragraph{Hitchin moduli space.} For our purposes it is better to consider the four dimensional theory on $S^1_R \times \real^3$. The moduli space of vacua is now the Hitchin moduli space $\cM_H$ \cite{Seiberg:1996nz}. Such a space is hyperK\"ahler and carries a family of complex structures $J_\zeta$ parametrized by $\zeta \in \PP^1$. For a certain value of $\zeta$ the Hitchin moduli space parametrizes Higgs bundles over $\cC$, and is a fibration by compact tori over $\cB$.

In \cite{Gaiotto:2008cd,Gaiotto:2009hg} a series of Darboux coordinates $\{ X_\gamma (u , \zeta) \}$ were constructed. Such coordinates depend on the BPS spectrum in a piecewise linear way. At BPS walls, where $Z_{\gamma'} / \zeta \in \real_-$ they transform as $X_\gamma \longrightarrow K_{\gamma'}^{\Omega (\gamma' , u)} (X_\gamma)$, where
\be \label{KSdiffeo}
K_{\gamma'} (X_\gamma) = X_\gamma \, \left( 1 - X_{\gamma'} \right)^{\langle \gamma , \gamma' \rangle}
\ee
is the Kontsevich-Soibelman symplectomorphism \cite{Kontsevich:2008fj}. Furthermore they obey the twisted group algebra relation $X_\gamma \, X_{\gamma'} = (-1)^{\langle \gamma , \gamma' \rangle} \, X_{\gamma + \gamma'}$. The IR vevs of line operators can be expressed as $J_\zeta$ holomorphic functions in these coordinates.

Conjecturally the Hitchin moduli space admits a quantization \cite{Cecotti:2010fi,Gaiotto:2010be} in terms of coordinates $\mathsf{X}_\gamma ( u , \zeta)$ such that
\be
\sfX_\gamma \, \sfX_{\gamma'} \, y^{-\langle \gamma , \gamma' \rangle} =  \sfX_{\gamma + \gamma'} \, .
\ee
We refer to this way of writing noncommutative coordinate as \textit{normal ordering} \cite{Cecotti:2010fi}. The quantum coordinates inherit the same piecewise behaviour as their classical limit. They jump across BPS walls, the jump being given by a quantum version of \eqref{KSdiffeo} obtained via conjugation by the quantum dilogarithm function.

\paragraph{UV line defects.} Theories of class $\cS$ admit BPS line operators \cite{Gaiotto:2010be,Moore:2015szp}. Quantum field theories contain far more information than what is captured by local correlators. Part of this structure, for example concerning dualities or non-perturbative effects, only emerges when considering defects. For a quantum field theory with gauge group $G$, a line defect is specified in the UV by Lie algebra data, a pair of electric and magnetic weights $(\lambda_e , \lambda_m) \in \Lambda_w \times \Lambda_{mw} / \mathfrak{w}$, consisting of an element of the weight lattices of $\frak{g}$ and its Langlands dual modulo the action of the Weyl group.

In the case of theories of class $\cS$ a line defect can be specified by a path $\wp$ on the UV curve $\cC$. This can be seen via geometric engineering, where line defects are described as boundaries of M2 branes ending on a configuration of M5 branes. Such a path can be closed and associated with an irreducible representation of $G$; more general configurations, connecting boundary components or punctures are in general allowed. The line operator is extended in the time direction in the four dimensional spacetime, or wrapping the $S^1_R$ when we consider the theory on $S^1_R \times \real^3$, and sits at a point in $\real^3$, for example the origin.

Such operators have very interesting algebraic properties. As they collide in $\real^3$ they form an OPE whose coefficients are vector spaces \cite{Gaiotto:2010be}. In particular such algebra is very similar to the multiplicative algebra of Wilson loop operators in Chern-Simons theory, with vector spaces playing the role of structure constant. This suggests that when $\wp$ is closed one could single out a $\real \times \cC$ factor from the full $\real^3 \times S^1_R \times \cC$ geometry and consider product loops of the form ${\rm pt} \times \wp$ with $\rm pt$ a point in $\real$ and $\wp$ a loop in $\cC$. It was proposed in \cite{Witten:2011zz} that such a setup should be related to a generalization of Khovanov homology for these loops.

In the simplest case of rank two theories, line defects associated with closed loops on $\cC$ can be classified in terms of a version of the Dehn-Thurston classification of isotopy classes of non-selfintersecting curves \cite{Drukker:2009tz}. The general idea is to go to a weak coupling limit, specified by a pants decomposition of the curve $\cC_{g,n}$, associated with separating curves $c_i^{(g,n)}$.  Then elements of $(\lambda_e , \lambda_m) \in \Lambda_w \times \Lambda_{mw} / \frak{w}$ are in one to one correspondence with geometric parameters, specifying the homotopy intersection number $\wp \, \#  \, c_i^{(g,n)}$ and with the twists of $c_i^{(g,n)}$.

\paragraph{IR line defects.} The Seiberg-Witten construction for four dimensional field theories is based on properly understanding the IR behaviour in the Coulomb branch. While the UV theory is naturally formulated in terms of local correlation functions, in the IR the structure of the moduli space of vacua is deeply tied to a BPS spectral problem: the wall-crossing behaviour of BPS states can be understood as a rule to glue together local patches of the moduli space in order to form a globally defined object \cite{Gaiotto:2008cd}. The same idea extends to line defects and relates the UV vacuum expectation value to the framed BPS spectral problem.

When we consider the theory on $\real^3 \times S^1_R$ with the defect wrapping the $S^1$, the vev of the line operator can be expanded \cite{Gaiotto:2010be,Moore:2015szp}
\be \label{vevL}
\langle L_{\wp,u,\zeta} \rangle = \sum_\gamma \ \underline{\overline{\Omega}} (L, \gamma , u , \zeta) \ X_\gamma (u , \zeta) \, ,
\ee
where $X_\gamma (u , \zeta)$ are the Darboux coordinates over the Hitchin moduli space. A line operator determines a specific $J_\zeta$-holomorphic function on the Hitchin moduli space. Often we will neglect to write down explicitly the dependence on $\zeta \in \PP^1$ and $u \in \cB$.

The coefficients $\underline{\overline{\Omega}} (L, \gamma , u , \zeta) $ are degeneracies of framed BPS states. For appropriate values of the moduli such BPS states can be thought of as BPS particles bound to an infinitely massive dyonic particle. The framed BPS degeneracies are a specialization of the protected spin character
\be
\underline{\overline{\Omega}} (L ,\gamma , u , \zeta ; y) :=  \Tr_{\cH_{L, \gamma , u , \zeta}} y^{2 J_3} (-y)^{2 I_3}
\ee
where $J_3$ is a generator of $\mathfrak{so} (3)$, the spatial rotation symmetry group, and $I_3$ a generator of the R-symmetry group $\mathfrak{su}(2)_R$. The protected spin  character determines a quantum line operator
\be
\mathscr{L}_\wp = \sum_\gamma \ \underline{\overline{\Omega}} (L , u , \gamma ; y) \ \mathsf{X}_\gamma
\ee
in terms of the quantum coordinates on the Hitchin moduli space $\mathsf{X}_{\gamma}$. For the purpose of this note we will consider the operators $\mathsf{X}_{\gamma}$ and their commutative limits as formal parameters.

Physically, an IR line operator can be thought of as an infinitely massive ``core'' dyonic particle and the framed BPS states can be interpreted as ordinary BPS particles bounded to the core particle \cite{Gaiotto:2010be,Cordova:2013bza}. This is the \textit{core charge} $\gamma_c$ and specifies the line defect at a certain point of the moduli space. 

The functions $\langle L_{\wp,u,\zeta} \rangle$  are continuous functions over the moduli space, in the sense that upon crossing a BPS wall the Darboux functions $X_\gamma$ transform with a Kontsevich-Soibelman diffeomorphism \eqref{KSdiffeo}  and the framed degeneracies $ \underline{\overline{\Omega}} (L, \gamma , u , \zeta)$ transform in such a way as to compensate the discontinuity. This discontinuity corresponds to the physical process in which a framed BPS bound state decays or is formed by emitting or capturing an ordinary BPS particle.

The functions  $X_\gamma$ have the asymptotic behavior $X_\gamma \sim c_\gamma \exp \pi \frac{R}{\zeta} Z_\gamma $ as $\zeta \rightarrow 0$, with $c_\gamma$ a constant \cite{Gaiotto:2010be}. A similar expansion holds for the functions \eqref{vevL}. When taking this limit the least dominant term corresponds to the state with the lowest energy in the framed BPS spectrum, as measured by the central charge at a point in the moduli space. The charge $\gamma_c$ corresponding to this term defines the core charge, at that point in the moduli space.

In the general this term can be identified with the term with the lowest degree of the right hand side of \eqref{vevL} interpreted as a Laurent polynomial (eventually by looking at the central charges if this is ambiguous). For example, when the line defect is labelled in the UV by an irreducible representation, the degree of this term corresponds to the lowest weight of the representation, expressed in a basis of charges of the lattice $\Gamma$. There is however an exception to this; crossing certain walls in the moduli space, called \textit{anti-walls} in \cite{Gaiotto:2010be}, two terms can exchange their roles. One can show that whenever this happens the basis of charges of $\Gamma$ undergoes a basis transformation, resulting in a new core charge, related to the old one by a quiver mutation \cite{Gaiotto:2010be,Cirafici:2017iju}. In this sense, while the value of the core charge and the specific form of the line defect \eqref{vevL} depend on the point of the moduli space, this information carries on unambiguously over all the moduli space. In this note we will work in a specific chamber, determined by the choice of a stability condition.

\paragraph{Categorification of link invariants.} In this paper we will consider line operators in class $\cS$ theories. It is however worth mentioning a different construction, introduced by Witten to understand certain aspects of Khovanov homology \cite{Witten:2011zz}. In that construction one starts with a knot or a link $L$ in a three manifold $M_3$. Vacuum expectation values of Wilson loops in $M_3$, as computed via Chern-Simons theory, determine topological invariants of knots \cite{Witten:1988hf}. A particular case is the Jones polynomial, a certain Laurent polynomial $J (q , L)$ which is a function of a variable $q$. To the knot or link $L$, Khovanov homology associates a bigraded complex $\mathsf{Kov} (L)= \oplus_{m,n} \mathsf{Kov}^{m,n} (L)$. The Jones polynomial can be recovered by taking the Euler characteristics
\be \label{Kov}
J (q , L) = \Tr_{\mathsf{Kov} (L)} \, (-1)^F \, q^P \, .
\ee
In \cite{Witten:2011zz} such a formula is reproduced from gauge theory, where $\mathsf{Kov}$ is replaced by a Hilbert space of BPS states of a higher dimensional theory, defined as the cohomology of an appropriate supercharge. The gradings $F$ and $P$ are then interpreted as an R-symmetry generator and as the instanton number in a gauge theory.

In the present paper we will attempt a different strategy to describe the framed BPS degeneracies. To begin with, note that the Hilbert space of framed BPS states is graded by the electro-magnetic charge
\be
\cH_{L , u } = \bigoplus_{\gamma \in \Gamma_L} \, \cH_{L , u , \gamma}
\ee
and depends non trivially on $u \in \cB$. Each $ \cH_{L , u , \gamma}$ is furthermore graded by the $\mathfrak{so}(3)$ quantum number $J_3$, $\cH_{L , u , \gamma}^{\bullet}$. In principle there is a further grading corresponding to $\mathfrak{su}(2)_R$. We do not expect this grading to play a role, due to the no-exotic conjecture, according to which physical states are all $\mathfrak{su}(2)_R$ singlets \cite{Chuang:2013wt,DelZotto:2014bga}. 

In our construction the role of $\mathsf{Kov}$ will be played by \eqref{introEC}, which has the form of an Euler characteristic of a complex of sheaves. While it is interesting to highlight this formal similarity, we will not pursue these ideas in this note and leave them for future investigations.

A very interesting prescription for the categorification of certain canonical basis associated to cluster varieties was  proposed in \cite{Allegretti:2018arr}. Line defects can be described in theories of class $\cS [A_1]$ in terms of  laminations over the curve $\cC$. The construction of  \cite{Allegretti:2018arr} takes certain laminations and writes the associated IR line operator as a sum of cluster variables, whose coefficients are graded dimensions of the singular cohomology of the quiver grassmannian associated with framed quivers. It would be very interesting to understand the relation between our construction and  \cite{Allegretti:2018arr}.

\subsection{Framed BPS quivers}

\paragraph{Quivers and representation theory.} In theories of class $\cS$ one can understand the BPS spectrum in certain chambers of the moduli space by studying the representation theory of certain quivers. A quiver is an oriented graph $Q = (Q_0 , Q_1)$ consisting of a set of nodes $Q_0$ and a set of arrows $Q_1$ connecting the nodes. Its algebra of paths $\complex Q$ is defined by concatenation of paths whenever possible. A superpotential is a function $\cW \, : \, Q_1 \longrightarrow \complex Q$ given by the sum of cyclic monomials. The formal derivative $\partial_a$ for $a \in Q_1$ acts non-trivially by cyclically permuting the elements of each monomial containing $a$ until $a$ is at the first position and then removes it. By taking the quotient with respect to the two sided ideal defined by the equations $\partial_a \cW = 0$ one defines the Jacobian algebra $\mathscr{J}_\cW = \complex Q / \langle \partial \, \cW \rangle$.

A representation of $Q$ consists in the assignment of a vector space $V_i$ for each node $i \in Q_0$ with morphisms $X_a \in \Hom (V_i , V_j)$ whenever $a \in Q_1$ connects the node $i$ with the node $j$. We require such morphisms to be compatible with the relations $\partial \, \cW = 0$. We denote by $\mathsf{rep} (Q , \cW)$ the corresponding category of representation of a quiver with potential. This category is equivalent to the category $\mathscr{J}_\cW - \mathsf{mod}$ of left modules over $\mathscr{J}_\cW$. 

The representation space is defined as 
\be
\mathsf{Rep}_{\mbf d} (Q) = \bigoplus_{a \, : \, i \rightarrow j} \Hom_\complex (V_i , V_j) \, .
\ee
Similarly we define the subscheme $\mathsf{Rep}_{\mbf d} (Q , \cW) $ by imposing the relations $\partial \, \cW = 0$. In physical applications we always consider isomorphism classes of representations, under the action of the gauge group $G_{\mbf d} = \prod_{i \in Q_0} \, \mathrm{GL} (d_i , \complex)$. The relevant moduli stack of quiver representations is the quotient
\be
\cM_{\mbf d} (Q , \cW) = \mathsf{Rep}_{\mbf d} (Q , \cW) / G_{\mbf d} \, .
\ee

\paragraph{BPS quivers.} One constructs a BPS quiver in theories of class $\cS$ as follows \cite{Alim:2011ae,Alim:2011kw}. Assume $\{ \gamma_i \}$ is a positive basis of $\Gamma$ at $u \in \cB$, such that for each element $\gamma_i$ the central charge $Z_{\gamma_i} (u)$ lies in the upper half plane $\mathfrak{h}$. Then the nodes of the BPS quiver $Q$ correspond to the elements of the basis $\{ \gamma_i \}$, and two nodes $\gamma_i$ and $\gamma_j$ are joined by the signed number of arrows given by $\langle \gamma_i , \gamma_j \rangle$. 

Physical BPS states correspond to stable representations of $Q$. Elementary BPS constituents corresponding to the basis elements interact via the superpotential $\cW$. They can form bound states of charge $\gamma =  \sum_i \, d_i \, \gamma_i$ if a representation of $(Q , \cW)$ with dimensions $d_i = \dim \, V_i$ exists and is stable. Stability is an extra condition, determined by the central charge. The action of the central charge on the lattice of charges lifts to a map $Z (u) \, : \, K (\mathsf{rep} (Q , \cW)) \longrightarrow \complex$. A  state of charge $\gamma_r$ described by a representation $\mathsf{R} \in \mathsf{rep} (Q , \cW)$ is stable if for any proper sub-representation $\mathsf{S} \in \mathsf{rep} (Q , \cW)$ corresponding to a state of charge $\gamma_s$, we have that $\arg Z_{\gamma_s} (u) < \arg Z_{\gamma_r} (u)$.

At a point $u \in \cB$, assuming we can find a BPS quiver $Q$, the BPS spectral problem is now purely algebraic and consists in classifying all the stable representations of $Q$. This problem can be approached in a variety of methods \cite{Alim:2011ae,Alim:2011kw,Manschot:2010qz}.

\paragraph{Framed BPS quivers.} The formalism of BPS quivers can be generalized to include line defects. The low energy dynamics of a collection of BPS particles bound to an infinitely massive dyonic particle is captured by an effective quantum mechanics. Such model is obtained by considering the representation theory of a framed BPS quiver $Q[f]$. At a certain point $u \in \cB$, we add to the quiver $Q$ an extra node $f$ corresponding to the core charge $\gamma_f$ of the defect. This charge is an element of an extended lattice of charges $\Gamma_L$, which is a torsor for $\Gamma$. This node is connected to the rest of the quiver by using the symplectic pairing. Correspondingly we add a new term $\cW_L$ to the superpotential and we extend the central charge function to $Q[f]$ by linearity \cite{Cordova:2013bza}.

Determining $\cW_L$ can be in general difficult; a few examples where discussed in \cite{Cirafici:2017wlw}. The main difficulty is that when discussing a BPS quiver one has already taken a Wilsonian limit, by integrating out heavy degrees of freedom. As a consequence the coupling of the quiver to the framing node cannot in general be derived from known BPS quivers by sending the mass of a particle to infinity. Such a coupling can be determined by string theory engineering \cite{Chuang:2013wt} or by indirect methods \cite{Cirafici:2017wlw}. 

We define framed quiver representations as follows. Let us conventionally denote by $V_f$ the vector space based at the framing node. To model a line defect we require this vector space to be one dimensional, $V_f \simeq \complex$. Then the framed representation space is
\be
\mathsf{Rep}_{\mbf d} (Q [f]) = \mathsf{Rep}_{\mbf d} (Q) \oplus \bigoplus_{a : i \rightarrow f} \Hom_\complex (V_i , V_f) \oplus \bigoplus_{a : f \rightarrow i} \Hom_\complex (V_f , V_i) 
\ee
and $\mathsf{Rep}_{\mbf d} (Q [f] , \cW) $ refers to the sub-scheme cut out by the equations $\partial \, \cW = 0$, where now $\cW$ denotes the sum of the superpotential of the unframed quiver with $\cW_L$. This allows us to define the moduli space of framed representations
\be
\cM_{\mbf d} (Q[f] , \cW) = \mathsf{Rep}_{\mbf d} (Q[f] , \cW) / G_{\mbf d} \, .
\ee
Note that the gauge group $G_{\mbf d}$ does \textit{not} involve the framing node. Also note that a generic enough framing will break most or all the automorphisms of $\cM_{\mbf d} (Q , \cW)$, making $\cM_{\mbf d} (Q[f] , \cW)$ a much better behaved space.

Framed BPS quivers admit a particularly convenient choice of stability conditions. Cyclic stability conditions correspond to a physical situation where the phase of the central charge of the defect is much bigger than that of all the other particles involved \cite{szendroi}. This stability condition selects cyclic modules of $\mathscr{J}_\cW$, that is modules $M$ generated by a vector $v \in M$. Cyclic modules always arise as quotients of $\mathscr{J}_\cW$ by ideals, which makes them particularly useful in localization computations since one can use such ideals to parametrize fixed points, a fact we will use later. In particular our cyclic vectors $v$ will always be vectors of the form $C \, v_f$, where $v_f \in V_f$ and $C$ is a map from $V_f$ to one of the representation spaces $V$ of the quiver $Q$. Here we are using the physical requirement that $V_f \simeq \complex$. Therefore we will loosely talk of cyclic modules generated by $v_f \in V_f$. For a more in depth discussion we refer the reader to \cite{Cirafici:2017wlw,Cirafici:2018jor}.

Finally $\cM_{\mbf d} (Q[f] , \cW ; v)$ will denote the moduli space of cyclic modules generated by $v$. The main claim of \cite{Cirafici:2017wlw} is that Donaldson-Thomas invariants associated to these moduli spaces are the framed BPS degeneracies. This claim holds for a line defect modelled by a framing node $\gamma_f$ at a particular point in the moduli space where the BPS quiver description is valid. Virtual localization reduces the computation of these invariants to a combinatorial problem. In this paper we will generalize this statement by giving a prescription to compute the framed protected spin characters, by identifying them with refined Donaldson-Thomas invariants computed via localization.

\section{Quantum line operators from framed quivers} \label{Ktheory}

This Section contains our main construction, which determines quantum line operators in theories of class $\cS$ in terms of the K-theoretic Donaldson-Thomas theory of the moduli space $\cM_{\mbf d} (Q[f] , \cW ; v)$, under certain conditions.

\subsection{Framed BPS quivers from UV data}

To begin with one would like to derive a framed BPS quiver with superpotential starting from the UV data which specify a line defect. There is unfortunately no simple algorithmic way to do so, but a series of techniques which work in particular cases.

The simplest case is for $\cS [A_1]$ theories, which have rank 2. In this case on can derive the BPS quiver from an ideal triangulation $\cT$ on $\cC$ \cite{Alim:2011ae,Alim:2011kw}. An ideal triangulation $\cT$ is a collection of curves up to isotopy, which are mutually and self non-intersecting with the exception of the end points. Such curves can end at punctures or at the marked points and cannot be contracted to a boundary component or to a puncture. From such a triangulation one can associate a BPS quiver, whose nodes correspond to curves in the triangulation which are not boundary segments. The arrows between nodes are given by a pairing between the edges of the triangulation, while the superpotential is determined by a series of combinatorial rules \cite{Alim:2011ae,Alim:2011kw}. In the case where the line defect is specified by a loop on $\cC$ the core charge can be determined by knowing the dictionary between Lie algebra data and Dehn-Thurston like parameters, as explained in section \ref{generalities}. However one also needs to specify a superpotential involving the framing arrows, to be added to the superpotential of the BPS quiver. We are not aware of a systematic way of deriving this superpotential. Therefore we will only focus on the line defects for which this superpotential is known, for example from \cite{Cirafici:2017wlw}. 

Similar but less general arguments hold for higher rank theories. From the spectral network one can determine the BPS quiver in some region of the moduli space \cite{Gabella:2017hpz,Gang:2017ojg}. The core charge can in principle by determined by a careful study of the abelianiziation map for spectral networks \cite{Gaiotto:2012rg,Gabella:2016zxu} which relates flat connections on $\cC$ and on $\Sigma_u$. There is however no general algorithm to determine the superpotential $\cW_L$. A different possibility is to take a microscopic approach and determine the superpotential for the defect by counting holomorphic disks via the string engineering \cite{Chuang:2013wt,Eager:2016yxd}.

A particular case is when the line defect can also be described as a product loop of the form ${\rm pt} \times \wp$ on $\real \times \cC$. In this case we associate a K-theoretic enumerative problem to the geometric loop. We will see that the $\cN=(2,0)$ superconformal field theory provides a natural way to associate (equivariant) K-theory classes to such a loop. This prescription should be compared with \cite{Witten:2011zz}, and we hope to return to it in the future.

From now on we will assume that a pair $(\gamma_f , \cW_L)$ is given. Section \ref{computations} will contain a few examples. 

\subsection{K-theoretic framed BPS states}

We will now use the framed BPS quiver formalism to define a K-theoretic enumerative problem associated to a line defect. To our framed quiver we can associate the moduli space $\cM_{\mbf d} (Q[f] , \cW ; v)$ of stable quiver representations. To ease the notation we will denote this moduli space as $\mathscr{M}_\gamma$, using the relation between charges and dimensions explained in Section \ref{generalities}, and set $\mathscr{\widetilde{M}}_\gamma$ to be the free fields moduli space obtained without imposing the relations $\partial \, \cW = 0$. Furthermore we use the notation $\mathscr{M} = \sqcup_\gamma \mathscr{M}_\gamma$ and similarly for $\mathscr{\widetilde{M}}$.

We can regard the equations $\partial \, \cW = 0$ as defining a section $s$ of a certain bundle or sheaf $\mathscr{E}$ over $\mathscr{\widetilde{M}}$. $\mathscr{E}$ is known as the obstruction bundle and by definition
\be
\mathscr{M} = s^{-1} (0) \subset \mathscr{\widetilde{M}} \, .
\ee
A obvious enumerative problem is to compute $\chi (\mathscr{M} , \cO_{\mathscr{M}})$. However this is not natural from the perspective of supersymmetry \cite{Okounkov:2015spn}. Supersymmetry is associated with virtual counts, or indices, which require a certain grading. A grading appears naturally by considering a resolution $\mathscr{A}^\bullet$ of the sheaf $\cO_{\mathscr{\widetilde{M}}}$. In general $\mathscr{A}^\bullet$ is a sheaf of differential graded algebras. For example one can consider the Koszul resolution $(\bigwedge^\bullet \mathscr{E}^\vee , \dd)$, where $\dd$ is the contraction with the section $s$. Then $\cO_{\mathscr{M}}$ arises as the cohomology in degree zero of this resolution. In homological algebra it is not natural to consider only the zeroth cohomology, but one would rather consider the \textit{virtual structure sheaf} \cite{fantechi,Okounkov:2015spn}
\be
\cO_{\mathscr{M}}^{\rm vir} = \sum_i (-1)^i \mathscr{A}^i = \sum_i (-1)^i \, H^i (\mathscr{A}^\bullet) \, .
\ee
The computation of the numerical Euler characteristics $\chi^{\rm num} (\mathscr{M} ,\cO_{\mathscr{M}}^{\rm vir} )$ via equivariant localization corresponds to the problem of computing Donaldson-Thomas invariants, or framed BPS degeneracies in this context, and was solved in \cite{Chuang:2013wt,Cirafici:2017wlw} for a large class of quivers.

In this paper we will compute Euler characteristics as K-theory classes adapting the formalism developed in \cite{Nekrasov:2014nea} to study membranes on toric Calabi-Yaus, and in particular applied to the Hilbert scheme of points on $\complex^3$. One of the key points is that the relevant Euler characteristics will be valued in the \textit{modified} virtual structure sheaf. The appropriate object to consider is the virtual tangent space, which parameterizes obstructions and deformations
\be
\mathsf{T}^{\rm vir} = \mathrm{Def} - \mathrm{Obs}
\ee
where the sheaf parametrizing obstructions is determined by $\mathscr{E}$. Part of the problem is to determine $\mathsf{T}^{\rm vir}$ for each model, and we will see several examples in the next Section. Technically, in all the cases we will consider in this note, $\mathsf{T}^{\rm vir}$ is determined by an underlying obstruction theory (eventually by choosing $\cW_L$ appropriately). We stress that in this note we will only work equivariantly with respect to a certain toric action. Assuming we are given the virtual tangent space, we introduce
\be
\mathcal{K}^{1/2}_{\rm vir} = \mathrm{det}^{-1/2} \, \mathsf{T}^{\rm vir}
\ee 
and define the modified virtual structure sheaf
\be
\widehat{\cO}_{\mathscr{M}}^{\rm vir} \sim \cO_{\mathscr{M}}^{\rm vir} \otimes \mathcal{K}^{1/2}_{\rm vir} 
\ee
where by using the proportionality symbol we reserve the possibility of including counting parameters in the definition, which in our case will correspond to the $\sfX_\gamma$ variables.

We can now define more precisely our K-theoretic enumerative problem. We introduce K-theoretic line operators
\be \label{defK-EC}
\mathscr{L}_\wp =  \chi \left( \mathscr{M} ,\widehat{\cO}_{\mathscr{M}}^{\rm vir} \right) \equiv
 \sum_{\gamma} \, \chi \left( \mathscr{M}_\gamma , \cO_{\mathscr{M}_\gamma}^{\rm vir} \otimes \mathcal{K}^{1/2}_{\rm vir} \right) \, \mathsf{X}_\gamma 
\ee
where we consider the variables $\mathsf{X}_\gamma$ as formal counting parameters, introduced in the definition of $\widehat{\cO}_{\mathscr{M}}^{\rm vir}$. To avoid introducing unnecessary notation we use the same symbol $\sfX_\gamma$ for the formal counting parameters and for the quantum coordinates on the Hitchin moduli space. No ambiguity is possible as long as the physical coordinates are always normal ordered, in the sense of Section \ref{generalities}.

We stress that the Euler characteristics in \eqref{defK-EC} gives a K-theory class. We will now give a concrete prescription to evaluate $\mathscr{L}$ using localization in equivariant K-theory. Afterwards we will discuss how to extract protected spin characters from our definitions.

\subsection{Localization}

We can obtain more concrete expressions for the operators $\mathscr{L}_\wp$ by working equivariantly  with respect to a natural toric action. We assume we have a framed quiver $Q[f]$ with superpotential $\cW$, including the $\cW_L$ term. After imposing cyclic stability conditions, the relevant moduli space $\mathscr{M}$ parametrizes cyclic modules in the Jacobian algebra $\mathscr{J}_\cW$, generated by a certain vector $v \in V_f$ based at the framing node.

\paragraph{Toric action and fixed points.} The space $\mathscr{M}$ carries a natural toric action, obtained by rescaling by a factor $t_a$ every morphism $X_a \in \Hom_\complex (V_i , V_j)$ where $a : i \longrightarrow j$. In order to lift to $\mathscr{M}$ this action has to respect the equations $\partial \, \cW = 0$. This condition defines a sub-torus $\mathbb{T}_{F , \partial \cW}$ of $\mathbb{T}_F = \left( \complex^* \right)^{|Q_1 [f]|}$. On the other hand representations in $\mathscr{M}$ are defined up to isomorphisms and therefore we have to mod out by the sub-torus $\mathbb{T} = \left( \complex^* \right)^{|Q_0| - 1} \subset G_{\mbf d}$. The reason the exponent is $|Q_0|-1$ is that diagonal gauge transformations act trivially. Therefore the torus acting on $\mathscr{M}$ is $\mathbb{T}_{\cW} = \mathbb{T}_{F , \partial \cW} / \mathbb{T}_G$. Note that consistency requires each term in $\cW$ to carry the same toric weight. We will denote this weight by $\kappa$.

For the cases considered in this paper, fixed points of $\mathbb{T}_{\cW}$ have a combinatorial classification in terms of pyramid partitions. A pyramid partition $\pi = \{ \pi_i \}_i$ is a certain combinatorial object consisting of a collection of coloured stones, whose number characterizes a particular cyclic module by the identification $\pi_i = \dim_\complex V_i$. This procedure was discussed in detail in \cite{szendroi,Cirafici:2017wlw} and simplifies greatly the localization computation. However in this note we will only focus on simple cases to exemplify our formalism. For pedagogical purposes we will list directly the $\torus$-fixed modules. The more powerful techniques from \cite{Cirafici:2017wlw}  are however necessary for more complicated examples.

\paragraph{Virtual tangent space.}

Around a fixed point labelled by $\pi$ we can use the formalism of \cite{Cirafici:2017wlw} to write down the quiver deformation complex. In all the examples discussed in  \cite{Cirafici:2017wlw} and in this paper, it has the form 
\begin{equation}
\xymatrix@C=8mm{  0 \ar[r] & \mathsf{S}^0_\pi \ar[r]^{\delta_0} & \mathsf{S}^1_\pi \ar[r]^{\delta_1} & \mathsf{S}^2_\pi \ar[r]^{\delta_2} & \mathsf{S}^3_\pi \ar[r] & 0
} \, .
\end{equation}
Note that the class of quivers we are considering is not in general associated to Calabi-Yau algebras. The Calabi-Yau condition is a sufficient condition  to have a well defined enumerative problem, basically due to Serre duality. In the more general case, the condition to have a well posed problem reduces to checking that a certain pairing is determined by numerical invariants. This was done explicitly for BPS quivers of $SU(N)$ theories associated to toric threefolds via geometric engineering, including all their decoupling limits, in \cite{Chuang:2013wt}. This already covers a rather large class of BPS quivers. Similar arguments are expected to hold more in general, but in this note we will only consider this class.

At a fixed point $\pi$ we denote by $V_{i, \pi}$ the weight decomposition of the vector space $V_i$ as a $\mathbb{T}_{\cW} $-module. We then have
\begin{eqnarray}
\mathsf{S}^0_\pi &=& \bigoplus_{i \in Q_0} \, \mathrm{Hom}_\mathbb{C} (V_{i, \pi} , V_{i, \pi}) \, , \\
\mathsf{S}^1_\pi &=& \bigoplus_{\left( a : i \rightarrow j \right) \in Q_1 [f_{\mathbf{n}}]} \, \mathrm{Hom}_\mathbb{C}  (V_{i, \pi} , V_{j, \pi}) \otimes t_a \, ,\\
\mathsf{S}^2_\pi &=& \bigoplus_{\left( r_a : j \rightarrow k \right) \in \mathsf{R}} \, \mathrm{Hom}_\mathbb{C}  (V_{j, \pi} , V_{k, \pi}) \otimes \kappa \, t_a^{-1} \, =   \overline{ \mathsf{S}^1_\pi }  \otimes \kappa , \\
\mathsf{S}^3_\pi &=& \overline{\mathsf{S}^0_\pi}   \otimes \kappa \, .
\end{eqnarray}
The first term in the complex corresponds to infinitesimal gauge parameters and the map $\delta_0$ is the linearization of a gauge transformation. The term $\mathsf{S}^1_\pi$ corresponds to all the fields $X_a$ associated to the arrows of the framed quiver $a \in Q_1 [f]$. Each field carries a toric weight $t_a$ by definition. The map $\delta_1$ is a linearization of the relations $\partial \cW = 0$. The term $\mathsf{S}^2_\pi$ corresponds to the relations derived from the superpotential $r_a = \partial_a \, \cW$. Each summand in $\mathsf{S}^2_\pi$ is naturally dual to a term in $\mathsf{S}^1_\pi$, the field $X_a$ for which $r_a$ is the equation of motion. Note that since the superpotential $\cW$ carries weight $\kappa$, the toric weight of $r_a$ is $\kappa \, t_a^{-1}$. The map $\delta_2$ is associated to linearized relations between the relations, and the space $\mathsf{S}^3_\pi $ is the dual of $\mathsf{S}^0_\pi $ up to the weight of the superpotential.

Now we can write down the virtual tangent space at the fixed point $\pi$
\be
\mathsf{T}^{\rm vir}_\pi = \left( - \mathsf{S}^{0}_\pi + \mathsf{S}^1_\pi \right) - \kappa \left(-  \overline{\mathsf{S}^{0}_\pi} + \overline{\mathsf{S}^1_\pi} \right) = \sum_i w_i - \sum_i \frac{\kappa}{w_i} \, .
\ee
We have schematically denoted by $w_i$ the toric weights of each summand. Note that deformations and obstructions are exactly paired up to the weight $\kappa$.

Let us introduce Okounkov's function
\be
\hat{\mathsf{a}} (\mathsf{T}^{\rm vir}_\pi) = \prod_w \frac{(\kappa/w)^{1/2} - (w / \kappa)^{1/2}}{w^{1/2}-w^{-1/2}} \, .
\ee
Then in localized K-theory we have
\be
\widehat{\cO}_{\mathscr{M}_\gamma}^{\rm vir} = \sum_{\pi_i = d_i} \, \hat{\mathsf{a}} \left( \mathsf{T}^{\rm vir}_\pi \right) \, \cO_{\cI_\pi} \, ,
\ee
where $\pi_i = \dim_\complex V_i$ and $\cI_\pi$ is the ideal sheaf corresponding to $\pi$. The relation between charges and the dimension vector is as usual $\gamma = \gamma_f + \sum_i d_i \gamma_i$

\paragraph{K-theoretic and quantum line operators.} Finally all the ingredients are in place and we can write down an explicit form for our K-theoretic line operators
\begin{align}
\mathscr{L}_\wp & =  \chi \left( \mathscr{M} ,\widehat{\cO}_{\mathscr{M}}^{\rm vir} \right) =
 \sum_{\gamma} \, \chi \left( \mathscr{M}_\gamma , \cO_{\mathscr{M}}^{\rm vir} \otimes \mathcal{K}^{1/2}_{\rm vir} \right) \, \mathsf{X}_\gamma  =  \sum_{\gamma} \, \sum_{\pi_i = d_i} \, \hat{\mathsf{a}} \left( \mathsf{T}^{\rm vir}_\pi \right)  \, \mathsf{X}_\gamma \, .
\end{align}
We will discuss now how to extract from a K-theoretic line operator a quantum line operator, a generating function of protected spin characters. They key concept is \textit{rigidity} \cite{Nekrasov:2014nea}. Roughly speaking rigidity means that some equivariant quantity computed via localization turns out to be independent of some of the toric weights. This can be typically shown by studying various limits of the toric weights. From a physical perspective, if we think of the steps of the localization computation as involving a ratio of fermionic and bosonic determinants around a fixed point, rigidity results are equivalent to a cancellation between certain fermionic and bosonic degrees of freedom. We expect rigidity results to hold for highly symmetric theories.

Consider scaling away the weights $w^{\pm 1} \longrightarrow \infty$ in such a way that $\kappa$ remains constant. Such a scaling identifies a \textit{slope} $s$, a one-parameter subgroup of the toric group $\torus_\cW$.  Then the rational function 
\be
 \frac{(\kappa/w)^{1/2} - (w / \kappa)^{1/2}}{w^{1/2}-w^{-1/2}} 
\ee
is bounded and non zero and goes to $- \kappa^{\mp 1/2}$. Therefore 
\be
\hat{\mathsf{a}} (\mathsf{T}^{\rm vir}_\pi) \longrightarrow (- \kappa^{1/2})^{\rm Index_{\pi}}
\ee 
where
\be
{\rm Index_{\pi}} = \# \{ i \, \vert \, w_i \longrightarrow 0 \} - \# \{ i \, \vert \, w_i \longrightarrow \infty \}
\ee
depends explicitly on the fixed point.

This construction can be used to define K-theoretic and refined Donaldson-Thomas invariants for toric threefolds  \cite{Nekrasov:2014nea}, as well as certain generalizations \cite{Benini:2018hjy,Nekrasov:2017cih,Nekrasov:2018xsb}. When applied to our moduli spaces, it provides a mathematical description of the protected spin character, by identifying them with refined Donaldson-Thomas invariants of framed quiver representations. Indeed the construction of \cite{Nekrasov:2014nea} applies almost verbatim due to the fact that we are considering gauge theories which can be engineered in toric threefolds \cite{Chuang:2013wt}.

In the following we will identify the toric weight $ - \kappa^{1/2}$  with the refinement parameter $y$. We can give a heuristic argument as follows. The formalism of Nekrasov and Okounkov is grounded in the setting of topological strings on a toric variety. In  \cite{Nekrasov:2014nea} they show that after scaling away the toric weights the K-theoretic Donaldson-Thomas invariants reproduce the refined BPS invariants where the toric weight $ - \kappa^{1/2}$ is identified with the refinement parameter $y$. The main argument is that in a geometric setting the sheaf $\mathcal{K}^{1/2}_{\rm vir}$ plays the role of the square root of the virtual canonical bundle on the toric variety; tensoring by it as in \eqref{defK-EC} is responsible for passing from Dolbeaut to Dirac cohomology of the complex $\cO^{\rm vir}_\mathscr{M}$. In this sense the equivariant weight of $\mathcal{K}^{1/2}_{\rm vir}$ keeps track of spin information. A direct computation precisely identifies such equivariant weight as the counting parameter $y$ in the $\chi_{-y}$ genus of the moduli space of states when this is smooth; $\chi_{-y}$ can then be conjecturally identified with the refined index \cite{Chuang:2013wt}. In particular since the weight of the canonical bundle spans a one dimensional torus $\complex^*$, there are several ways to scale away the toric weights by keeping $\kappa$ constant. Such possibilities are parametrized by the slope parameter $s$, which therefore parametrizes different conventions in the choice of the refinement.

While we are not in a geometric setting we would like to argue that such arguments should extend to our case upon using geometric engineering to describe our quiver quantum mechanics as a limit of the quantum mechanics modelling a D-brane configuration. Such a field theory limit would take a D-brane configuration wrapping certain cycles in a Calabi-Yau and tune their moduli in such a way that string states decouple, while keeping the mass of one field theory state much larger than all the others, such state modelling the defect. However by taking such limit we lose the intuition coming from geometrical engineering and the dictionary between toric parameters and the threefold geometry is no longer available. Still we conjecture that the combinations of the toric weights which survives the scaling limit $w^{\pm} \longrightarrow \infty$ along a certain slope $s$, while it has no more the interpretation as the weight of the Calabi-Yau canonical bundle, can still be identified with the refined parameter $y$.

If we identify $y = - \kappa^{1/2}$ and use the slope $s$ to denote that we have taken the appropriate limit and scaled away the toric weights, we have 
\be \label{qL-index}
\mathscr{L}_\wp \big\vert_s = \sum_{\gamma} \, \left( \Tr_{\cH_{L, u ,\gamma}} y^{2 J_3} (-y)^{2 I_3} \right) \ \mathsf{X}_\gamma = \sum_{\gamma} \left( \sum_{\pi \in \mathscr{M}_\gamma} y^{\rm Index_{\pi}} \right) \ \mathsf{X}_\gamma \, .
\ee
This generating function has very interesting properties in the case where $\mathscr{M}_\gamma$ is compact. If that is the case, then  \cite{Nekrasov:2014nea} shows that the toric action factors throught the character $\kappa$: the result only depends on $\kappa$ and no limit is needed. This also implies that \eqref{qL-index} is a Laurent polynomial, invariant under $y \longleftrightarrow 1/y$. In general we do not know if our moduli space $\mathscr{M}_\gamma$ are compact for any pair $(\sfQ [f], \cW)$ which arises from class $\cS$ theories. In all examples we have found it appears to be the case and in all the cases discussed in Section \ref{computations} the results are independent of the toric weights without taking any limit. For this reason we will not stress the difference between $\mathscr{L}_\wp$ and $ \mathscr{L}_\wp \big\vert_s$ in Section \ref{computations}.  It is not clear if this is a general feature of line defects or if the quivers we are considering are too simple. For example the results are explicitly dependent on the toric weights in the more complicated case of Calabi-Yau singularities \cite{MQ}.

\section{Computations with quantum line operators} \label{computations}

In this Section we will go through some detailed computations to exemplify our formalism. We will also clarify the reason for using certain shifted superpotentials associated with symmetric quivers \cite{Cirafici:2017wlw} and the difficulties which one runs into when dealing with localization with a system with nontrivial automorphisms.

\subsection{Quiver for SU(2) with adjoint line defect}

In our first example we consider an SU(2) gauge theory coupled to a Wilson line in the adjoint representation. This model was already studied in \cite{Cirafici:2017wlw}, and we will freely borrow the details non essential to the present construction.

\paragraph{UV data.} To obtain a pure SU(2) gauge theory from class $\cS$ engineering, $\cC$ must be an annulus with a marked point at each boundary. The ideal triangulation $\cT$ has two edges. We consider a loop $\wp$ in $\cC$ wrapping the annulus once, in the adjoint representation $\mathbf{3}$ of SU(2).

\paragraph{Framed BPS quiver.} This line defect has core charge  $\gamma_f = - \gamma_\circ - \gamma_\bullet$, corresponding to the highest weight of the $\mbf 3$ of SU(2) expressed in the quiver basis of charges $\{ \gamma_\circ , \gamma_\bullet \} \in \Gamma$. The quiver corresponding to the triangulation $\cT$ is the 2-Kronecker quiver. The two charges associated to the two edges of the triangulation reproduce the BPS spectrum in the strong coupling chamber. The framed BPS quiver is

\begin{equation}
\xymatrix@C=8mm{
& \gamma_f   \ar@{..>}@<-0.5ex>[ddl]_{C_1}  \ar@{..>}@<0.5ex>[ddl]^{C_2}  & \\
 & & \\
 \gamma_\circ   \ar@<-0.5ex>[rr]_{A_1}  \ar@<0.5ex>[rr]^{A_2} & & \gamma_\bullet   \ar@{..>}@<-0.5ex>[uul]_{B_1}  \ar@{..>}@<0.5ex>[uul]^{B_2} 
}
\end{equation}

The gauge group $G_{\mbf d} = \mathrm{GL} (V_\circ) \times \mathrm{GL} (V_\bullet)$ acts as $A_i \longrightarrow g_\bullet \, A_i \, g_\circ^{-1}$, $B_i \longrightarrow B_i \, g_\bullet^{-1}$ and $C_i \longrightarrow g_\circ \, C_i$, for $i=1,2$. Following \cite{Cirafici:2017wlw} we take the superpotential
\begin{equation} \label{SU2adjW}
\mathcal{W} = A_1 \, C_1 \, B_1 + B_2 \, \left( A_2 \, C_2- A_2 \, C_1 \right) \, .
\end{equation}
For now we take this superpotential as given; we will justify its form later on. The equations of motions $\partial_a \, \cW = 0$ can be derived easily and also satisfy the following relations between the relations
\begin{align}
rr_{\bullet} \, : \qquad  A_1 \, \partial_{A_1} \cW - \partial_{B_1} \cW \, B_1 - \partial_{B_2} \cW \, B_2 + A_2 \, \partial_{A_2} \cW &= 0 \, , \\
rr_{\circ} \, : \qquad  C_1  \, \partial_{C_1} \cW - \partial_{A_1} \cW \, A_1 -\partial_{A_2} \cW \, A_2 + C_2 \, \partial_{C_2}  \cW &= 0 \, . 
\end{align}

\paragraph{Toric action and fixed points.} We define the toric action $X_a \longrightarrow t_{X_a} \, X_a $ for each morphism $X_a$ associated with an arrow $a \, : \, i \longrightarrow j$, a field in the supersymmetric quiver quantum mechanics. We require that this action is compatible with the equations of motion, since these define the moduli space. From the equations of motion we have the identifications
\be \label{3SU2-toric-cond}
t_{C_1} = t_{C_2} \, \qquad t_{A_1} \, t_{B_1} = t_{A_2} \, t_{B_2} \, .
\ee
The superpotential carries weight $t_{A_1} t_{B_1} t_{C_1} = t_{A_2} t_{B_2} t_{C_2} = \mathbf{\kappa}$.

The $\torus$-fixed points are given by $\torus$-fixed cyclic modules, corresponding to $\torus$-fixed ideals in the path algebra. They were classified in \cite{Cirafici:2017wlw} and can be counted combinatorially by listing pyramid partitions. However, since they are just a few, we will list them explicitly to keep this example as simple as possible. To begin with, consider the set of vectors
\be
\{ v , C_1 \, v , C_2 \, v , A_2 \, C_1 \, v , A_1 \, C_2 \, v , \dots \}
\ee
where we have used the $\partial \, \cW = 0$ relations to set $A_1 \, C_1 \, v = 0$ and $A_2  \, C_2 \, v = A_2 \, C_1 \, v$. These vectors are obtained by applying arrows to the cyclic vector and imposing the F-term relations at each step. They are therefore linearly independent. Therefore we can use them to identify cyclic modules. We assign the vectors to their respective nodes of the quiver. Since they are linearly independent they generate vector spaces. The action of the elements of the path algebra on these vectors can be used to assign maps between the vector spaces they generate. In other words we are constructing a cyclic representation, starting from $v$. Furthermore each map has its own toric weight, which can be compensated by a gauge transformation at each vertex. Therefore these modules are $\torus$-fixed. Physically we want to impose the condition that $\dim_\complex V_f = 1$, which means that we only restrict our attention to those representations which have a single vector $v \in V_f$. For example we cannot have vectors of the form $B_i A_1 C_2 v$ : such vector is based at the framing node and is linearly independent from $v$, having different toric weight. This is not possible because of the condition $\dim_\complex V_f = 1$. Therefore the arrows  $B_i$ effectively play no role. 

Fixed points are associated with the cyclic representations, which in turns are determined by the above vectors. We will use the notation $\{ \dots \}_{1,d_\circ,d_\bullet}$ to identify a fixed point together with its dimension. With this notation the fixed points are
\begin{gather}
\{ v  \}_{1,0,0} , \{ v , C_1 v \}_{1,1,0} , \{ v , C_2 v \}_{1,1,0} , \{ v , C_1 v  , C_2 v \}_{1,2,0} ,  \{ v , C_2 \, v , A_1 \, C_2 v \}_{1,1,1} , \\
  \{v , C_1 v , A_2 C_1 v , C_2 v \} _{1 ,2 ,1}, \{ v , C_1 v ,C_2 v ,  A_1 C_2 v \}_{1,2,1} , \{ v , C_1 v , C_2 v , A_2 C_1 v , A_1 C_2 v \}_{1,2,2} 
\end{gather}
where again we have used $A_1 \, C_1 \, v = 0$ and $A_2  \, C_2 \, v = A_2 \, C_1 \, v$. This completes the classification of $\torus$-fixed points. A more detailed discussion is in \cite{Cirafici:2017wlw}.

\paragraph{Virtual tangent space.} We can construct the virtual tangent space as explained in Section \ref{Ktheory} starting from the deformation complex
\begin{equation}
\xymatrix@C=8mm{  0 \ar[r] & \mathsf{S}^0_\pi \ar[r]^{\delta_0} & \mathsf{S}^1_\pi \ar[r]^{\delta_1} & \mathsf{S}^2_\pi \ar[r]^{\delta_2} & \mathsf{S}^3_\pi \ar[r] & 0 \, ,
}
\end{equation}
where $\delta_0$, $\delta_1$ and $\delta_2$ are linearization of the gauge transformations, equations of motion, and relations between the relations (around a fixed point). Their explicit form is given in \cite{Cirafici:2017wlw}.

The virtual tangent space parametrizes deformations minus obstructions and in this case has the form
\begin{equation} \label{Tvir3SU2}
\mathsf{T}^{\rm vir}_\pi = - \mathsf{S}^0_\pi + \mathsf{S}^1_\pi - \mathsf{S}^2_\pi + \mathsf{S}^3_\pi 
= - \mathsf{S}^0_\pi + \mathsf{S}^1_\pi - \mathbf{\kappa} \left(- \overline{\mathsf{S}^0_\pi} + \overline{\mathsf{S}^1_\pi} \right) = \sum_i w_i - \sum_i \frac{\kappa}{w_i} \, .
\end{equation}
In writing this formula one has to take into account the decomposition of the vector spaces $V_{\circ,_\pi}$ and $V_{\bullet,_\pi}$ as $\torus$-modules. The terms in \eqref{Tvir3SU2} are as follows:
\be
\mathsf{S}^0_\pi = \Hom_\complex (V_{\circ,_\pi} , V_{\circ,_\pi}) \oplus \Hom_\complex (V_{\bullet,_\pi} , V_{\bullet,_\pi})
\ee
parametrizes gauge transformations (note that there are no gauge transformations at the framing vertex),
\begin{gather}
\mathsf{S}^1_\pi = \Hom_\complex (V_{\circ,_\pi} , V_{\bullet,_\pi}) (t_{A_1} + t_{A_2}) \oplus \Hom_\complex (V_{\bullet,_\pi} , \complex ) (t_{B_1} + t_{B_2}) \oplus \Hom_\complex (\complex , V_{\circ,_\pi}) (t_{C_1} + t_{C_2})
\end{gather}
parametrizes the field content (where the toric weights are needed to make the deformation complex equivariant), while
\begin{align}
\mathsf{S}^2_\pi &= \Hom_\complex (V_{\bullet,_\pi} , V_{\circ,_\pi}) (t_{B_1} t_{C_1} + t_{B_2} t_{C_2}) \oplus \Hom_\complex (\complex , V_{\bullet,_\pi}) (t_{A_1} t_{C_1} + t_{A_2} t_{C_2})  
\cr  
&\oplus \Hom_\complex (V_{\circ,_\pi} , \complex) (t_{A_1} t_{B_1} + t_{A_2} t_{B_2})  
\cr
&= \mathbf{\kappa} \, \Big[ 
 \Hom_\complex (V_{\bullet,_\pi} , V_{\circ,_\pi}) (t_{A_1}^{-1} + t_{A_2}^{-1} ) \oplus \Hom_\complex (\complex , V_{\bullet,_\pi}) (t_{B_1}^{-1} + t_{B_2}^{-1})  
 \cr  
 &\oplus \Hom_\complex (V_{\circ,_\pi} , \complex) (t_{C_1}^{-1} + t_{C_2}^{-1} ) 
\Big] = \mathbf{\kappa} \, \overline{\mathsf{S}^1_\pi}
\end{align}
parametrizes the equations of motion (where we have used $\mathbf{\kappa} = t_{A_1} t_{B_1} t_{C_1} = t_{A_2} t_{B_2} t_{C_2} $). Finally
\be
\mathsf{S}^3_\pi =  \Hom_\complex (V_{\circ,_\pi} , V_{\circ,_\pi}) \mathbf{\kappa}  \oplus \Hom_\complex (V_{\bullet,_\pi} , V_{\bullet,_\pi}) \mathbf{\kappa} = \mathbf{\kappa} \ \overline{\mathsf{S}^0_\pi}
\ee
parametrizes doubly determined relations.

\paragraph{Contribution of fixed points.} Now we will carry out the localization computation in a fairly explicit fashion. We organize the fixed point contribution according to the dimension vector $\mathbf{d} = (1 , d_\circ , d_\bullet)$ of the representation. Each fixed point with $\mathsf{T}^{\rm vir}_\pi = \sum_i w_i - \sum_i \kappa / w_i$ contributes with a factor $\hat{\mathsf{a}}  ( \sum_i w_i - \sum_i \kappa / w_i) $.
\begin{enumerate}
\item 
\underline{$\mathbf{d} = (1,0,0)$}. In this case we simply have
\begin{gather}
V_{\circ,\pi} = 0 \, , \qquad V_{\bullet,\pi} = 0 \, ,
\end{gather}
and $\mathsf{T}_\pi^{\rm vir} = 0$. Therefore the contribution of this fixed point is $1$.

\item
\underline{$\mathbf{d} = (1,1,0)$}. There are two fixed points, which we will call $\pi_1$ and $\pi_2$, corresponding to $\{ C_1 v \} $  and $ \{ C_2 v \}$. The $\torus$-module structure associated to the first one is
\be
V_\circ = \frac{1}{t_{C_1}} \, , \qquad V_\bullet = 0 \, .
\ee
To compute the weight decomposition of the virtual tangent space we need to specify the spaces $\mathsf{S}_\pi^0$ and $\mathsf{S}_\pi^1$. We find, writing explicitly only the non-vanishing part
\begin{align}
\mathsf{S}_\pi^0 &= \Hom_\complex (V_{\circ,_\pi} , V_{\circ,_\pi}) = V_{\circ,_\pi} \otimes V_{\circ,_\pi}^\vee = 1 \, , \cr
\mathsf{S}_\pi^1 &= \Hom_\complex (\complex , V_{\circ,_\pi}) (t_{C_1} + t_{C_2}) = \frac{1}{t_{C_1}} (t_{C_1} + t_{C_2}) = 1 + \frac{t_{C_2}}{t_{C_1}} \, . 
\end{align}
Therefore $-\mathsf{S}_\pi^0 +\mathsf{S}_\pi^1 =  \frac{t_{C_2}}{t_{C_1}}$ and the virtual tangent space reads
\be
\mathsf{T}_{\pi_1}^{\rm vir}= \frac{t_{C_2}}{t_{C_1}} - \mathbf{\kappa} \frac{t_{C_1}}{t_{C_2}} \, .
\ee
For the second fixed point we have
\be
V_\circ = \frac{1}{t_{C_2}} \, , \qquad V_\bullet = 0 \, .
\ee
Reasoning as before we find
\be
\mathsf{T}_{\pi_2}^{\rm vir} = \frac{t_{C_1}}{t_{C_2}} - \mathbf{\kappa} \frac{t_{C_2}}{t_{C_1}} \, .
\ee
Putting all together the two fixed points contribute 
\be
\hat{\mathsf{a}} \left(\mathsf{T}_{\pi_1}^{\rm vir} \right)  + \hat{\mathsf{a}} \left( \mathsf{T}_{\pi_2}^{\rm vir} \right)  = 
 \frac{-\mathbf{\kappa}\,  t_{C_1} + t_{C_2}}{\sqrt{\mathbf{\kappa}} \, (t_{C_1} - t_{C_2})}
 +
 \frac{-\mathbf{\kappa}\,  t_{C_2} + t_{C_1}}{\sqrt{\mathbf{\kappa}} \, (t_{C_2} - t_{C_1})} 
 =
-\frac{1}{\sqrt{\kappa}} - \sqrt{\kappa} \, ,
\ee
which is independent of the toric weights.

\item
\underline{$\mathbf{d} = (1,2,0)$}. There is only one fixed point, corresponding to $ \{v , C_1 v  , C_2 v \} $, so that
\be
V_\circ = \frac{1}{t_{C_1}} + \frac{1}{t_{C_2}} \, , \qquad V_\bullet = 0 \, ,
\ee
and $\mathsf{T}_\pi^{\rm vir}=0$. The contribution of this fixed point is one.

\item
\underline{$\mathbf{d} = (1,1,1)$}. There is only one fixed point, with
\be
V_\circ = \frac{1}{t_{C_2}}  \, , \qquad V_\bullet = \frac{1}{t_{C_2} t_{A_1}} \, .
\ee
The virtual tangent space is given by
\be
\mathsf{T}_{\pi}^{\rm vir} 
   = \frac{t_{A_2}}{t_{A_1}}+t_{A_1} t_{B_1}
   t_{C_2}+t_{A_1} t_{B_2} t_{C_2}+\frac{t_{C_1}}{t_{C_2}} - \kappa \left(
   \frac{ t_{A_1}}{t_{A_2}}+\frac{1}{t_{A_1} t_{B_1}
   t_{C_2}}+\frac{1}{t_{A_1} t_{B_2} t_{C_2}}+\frac{
   t_{C_2}}{t_{C_1}}
    \right) \, .
\ee
Now we impose the conditions \eqref{3SU2-toric-cond} as well as $t_{A_1} t_{B_1} t_{C_1} = t_{A_2} t_{B_2} t_{C_2} = \mathbf{\kappa}$, and easily find that $\mathsf{T}_{\pi}^{\rm vir} = 0$ identically. Therefore the contribution of this fixed point is just one, a hypermultiplet.

\item
\underline{$\mathbf{d} = (1,2,1)$}. Now there are two fixed points. The first one $\pi_1$ corresponds to the module $ \{v , C_1 v , A_2 C_1 v , C_2 v \}$ and its $\torus$-module structure is given by 
\be
V_\circ = \frac{1}{t_{C_1}} + \frac{1}{t_{C_2}}    \, , \qquad V_\bullet = \frac{1}{t_{A_2} t_{C_1}}
\ee
The virtual tangent space is then
\begin{align}
\mathsf{T}_{\pi_1}^{\rm vir}  = & \  \frac{t_{A_1} t_{C_2}}{t_{A_2}
   t_{C_1}}+\frac{t_{A_1}}{t_{A_2}}+t_{A_2} t_{B_1}
   t_{C_1}+t_{A_2} t_{B_2} t_{C_1}+\frac{t_{C_2}}{t_{C_1}} \cr
   & - \kappa \left(
   \frac{ t_{A_2} t_{C_1}}{t_{A_1} t_{C_2}}
   + \frac{t_{A_2}}{t_{A_1}} + \frac{1}{t_{A_2} t_{B_1}
   t_{C_1}} + \frac{1}{t_{A_2} t_{B_2} t_{C_1}} + \frac{t_{C_1}}{t_{C_2}}
   \right) \cr
   = & \ \frac{t_{A_1}}{t_{A_2}} - \kappa \frac{t_{A_2}}{t_{A_1}}  \, ,
\end{align}
where we have used the conditions on the toric weights. Note that upon imposing these conditions there are many cancellations, which signal that writing down naively the virtual tangent space miscounts deformations and obstructions.

The second fixed point $\pi_2$ corresponds to the module $ \{ v , C_1 v ,C_2 v ,  A_1 C_2 v \}$, for which
\be
V_\circ = \frac{1}{t_{C_1}} +  \frac{1}{t_{C_2}}    \, , \qquad V_\bullet =\frac{1}{t_{A_1} t_{C_2}} \, .
\ee
The virtual tangent space can be simplified using again the conditions on the toric weights
\begin{align}
\mathsf{T}_{\pi_2}^{\rm vir}  = & \ 
   \frac{t_{A_2} t_{C_1}}{t_{A_1}
   t_{C_2}}+\frac{t_{A_2}}{t_{A_1}}+t_{A_1} t_{B_1}
   t_{C_2}+t_{A_1} t_{B_2} t_{C_2}+\frac{t_{C_1}}{t_{C_2}} \cr
   & \ - \kappa \left(
   \frac{t_{A_1} t_{C_2}}{t_{A_2} t_{C_1}}+\frac{
   t_{A_1}}{t_{A_2}}+\frac{1}{t_{A_1} t_{B_1}
   t_{C_2}}+\frac{1}{t_{A_1} t_{B_2} t_{C_2}}+\frac{
   t_{C_2}}{t_{C_1}}
   \right)
 \cr = & \ \frac{t_{A_2}}{t_{A_1}} - \kappa \frac{t_{A_1}}{t_{A_2}}  \, .
\end{align}
Putting the two contributions together gives
\be
\hat{\mathsf{a}} \left(  \frac{t_{A_1}}{t_{A_2}} - \kappa \frac{t_{A_2}}{t_{A_1}} 
\right)  + \hat{\mathsf{a}} \left(  \frac{t_{A_2}}{t_{A_1}} - \kappa \frac{t_{A_1}}{t_{A_2}} \right)  = -\frac{1}{\sqrt{\kappa}} - \sqrt{\kappa} \, .
\ee

\item
\underline{$\mathbf{d} = (1,2,2)$}. There is only one fixed point, with
\be
V_\circ = \frac{1}{t_{C_1}} +  \frac{1}{t_{C_2}}    \, , \qquad V_\bullet = \frac{1}{t_{A_2} t_{C_1}}  + \frac{1}{t_{A_1} t_{C_2}} \, ,
\ee
so that 
\begin{align}
\mathsf{T}_{\pi}^{\rm vir} = & \
   \frac{t_{A_2}}{t_{A_1}}+\frac{t_{A_1}}{t_{A_2}}+
   t_{A_1} t_{B_1} t_{C_2}+t_{A_1} t_{B_2} t_{C_2}+t_{A_2}
   t_{B_1} t_{C_1}+t_{A_2} t_{B_2}
   t_{C_1}+\frac{t_{C_2}}{t_{C_1}}+\frac{t_{C_1}}{t_{C_2}} \\  \nonumber
& - \kappa \left(
\frac{t_{A_2}}{t_{A_1}}+\frac{t_{A_1}}{t_{A_2}}+\frac{1}{t_{A_1}
   t_{B_1} t_{C_2}}+\frac{1}{t_{A_1} t_{B_2}
   t_{C_2}}+\frac{1}{t_{A_2} t_{B_1} t_{C_1}}+\frac{1}{t_{A_2}
   t_{B_2} t_{C_1}}
   +\frac{t_{C_2}}{t_{C_1}}+\frac{
   t_{C_1}}{t_{C_2}} 
\right) \, .
\end{align}
By imposing the conditions on the toric weights one sees that $\mathsf{T}_{\pi}^{\rm vir} = 0$ and therefore the contribution of this fixed point is just $1$. 

\end{enumerate}

\paragraph{Full result.}

The full result for the quantum line operator is
\begin{align}
\mathscr{W}_{\mathbf{3}}  = \sum_{d = (1 , d_\circ , d_\bullet)} 
\chi (\mathscr{M}_\gamma , \cO^{\rm vir} \otimes \mathcal{K}^{1/2}_{\rm vir})
\ \mathsf{X}_{\gamma_f  + d_\circ \gamma_\circ + d_\bullet \gamma_\bullet} \, .
\end{align}
As we have seen in this computation the dependence of all the individual toric weights drops out and only $\kappa$ remains. Therefore in this case we don't need to take any limit. The formal counting variables $\sfX_{\gamma}$ have to be identified with the quantum coordinates on the Hitchin moduli space. To avoid ambiguities we will always assume them normal ordered. It is important to stress that here $\chi (\mathscr{M}_\gamma , \cO^{\rm vir} \otimes \mathcal{K}^{1/2}_{\rm vir})$ is not a number but an equivariant K-theory class in $K_\torus ({\rm pt}) [[ \sqrt{k} ]]$. We can think of this as a vector space, or as the $\torus$-module generated by $\sqrt{\kappa}$.

By defining $y = - \sqrt{\mathbf{\kappa}}$ we can write the full result as
\begin{align}
\mathscr{W}_{\mathbf{3}}  &= \mathsf{X}_{\gamma_f} + \left( y + \frac{1}{y} \right) \mathsf{X}_{\gamma_f +  \gamma_\circ }  + \mathsf{X}_{\gamma_f + 2 \gamma_\circ} \\
&+\mathsf{X}_{\gamma_f + \gamma_\circ + \gamma_\bullet} 
+ \left( y + \frac{1}{y} \right) \ \mathsf{X}_{\gamma_f + 2 \gamma_\circ +  \gamma_\bullet}  + \mathsf{X}_{\gamma_f + 2 \gamma_\circ + 2 \gamma_\bullet}
\end{align}

The framed BPS spectrum consists of four hypermultiplets with charges $\gamma_f$ (the core charge), $\gamma_f + 2 \gamma_\circ$, $\gamma_f + \gamma_\circ+\gamma_\bullet$ and $\gamma_f + 2 \gamma_\circ + 2 \gamma_\bullet$, as well as two vector multiplets with charges $\gamma_f + \gamma_\circ$ and $\gamma_f + 2 \gamma_\circ + \gamma_\bullet$. This prediction is indeed correct and can be checked independently, for example using the fact that Wilson line operators obey an algebra, in this case derived from the tensor product decomposition of $SU(2)$ representations \cite{Gaiotto:2010be,Cordova:2013bza}.

\subsection{Localization for theories with automorphisms}

We will now discuss some remarks concerning the case where the BPS quiver has nontrivial automorphisms, in passing clarifying certain aspects of \cite{Cirafici:2017wlw}. In this case the direct localization computation has to be handled with care. The reason is that the presence of a nontrivial automorphism at the level of the quiver induces a discrete symmetry in the supersymmetric quantum mechanics. When this happens, one has to decide if the quantum mechanics particles are indistinguishable, and in that case their statistics \cite{Cecotti:2012va}. In such a case the naive representation theory result differs from the physical one, since the quantum mechanics models requires us to take linear combinations of the wavefunctions with prescribed symmetry, to construct physically acceptable ground states.

We will consider automorphisms which exchange certain arrows of the framed quiver but leave the nodes and the superpotential invariant. In this case the so-called $ii$-representations are involved \cite{Cecotti:2012va}. The problem of properly dealing with localization in the presence of $ii$-representations is the reason we have introduced shifted superpotentials (in the previous localization computation as well as in \cite{Cirafici:2017wlw}): in many cases a variable shift is enough to break the discrete symmetry at the level of the superpotential. This is the reason behind the superpotential \eqref{SU2adjW}, which we preferred to the more natural \eqref{SU2-3-symmW}. This does not have any physical effect, since it is just a field redefinition, but has the technical advantage of rendering the localization computation direct. 

We will discuss the situation with an example. Consider the framed BPS quiver $\sfQ$ which describes the coupling of an SU(2) gauge theory to a Wilson line in the adjoint representation
\begin{equation}
\xymatrix@C=8mm{
& \gamma_f   \ar@{..>}@<-0.5ex>[ddl]_{C_1}  \ar@{..>}@<0.5ex>[ddl]^{C_2}  & \\
 & & \\
 \gamma_\circ   \ar@<-0.5ex>[rr]_{A_1}  \ar@<0.5ex>[rr]^{A_2} & & \gamma_\bullet   \ar@{..>}@<-0.5ex>[uul]_{B_1}  \ar@{..>}@<0.5ex>[uul]^{B_2} 
}
\end{equation}
where now we take the superpotential
\begin{equation} \label{SU2-3-symmW}
\sfW = A_1 C_1 B_1 + A_2 C_2 B_2
\end{equation}
This theory has a $\zed_2$ symmetry which exchanges $X_1 \longleftrightarrow X_2$ for any field $X_i$, keeping the superpotential invariant. We would like to understand how to count BPS states using the localization framework that we have introduced so far. To begin with, we work equivariantly by introducing a toric action which rescale each field $X$ by a factor $t_X$. The condition that the toric action commutes with the $\zed_2$ symmetry implies $t_{X_1} = t_{X_2}$ for any field. We set $\kappa = t_{A_1} t_{B_1} t_{C_1} = t_{A_2} t_{B_2} t_{C_2}$, the weight of the superpotential.

The setup for the localization computation proceeds as before, the only difference being the superpotential equations $C_1 \, A_1 = C_2 \, A_2 = 0$. The $\torus$-fixed modules are now
\begin{gather}
\{ v  \}_{1,0,0} , \{ v , C_1 v \}_{1,1,0} , \{ v , C_2 v \}_{1,1,0} , \{ v , C_1 v  , C_2 v \}_{1,2,0} ,  \{ v , C_2 \, v , A_1 \, C_2 v \}_{1,1,1} ,  \{ v , C_1 \, v , A_2 \, C_1 v \}_{1,1,1} \cr
  \{v , C_1 v , A_2 C_1 v , C_2 v \} _{1 ,2 ,1}, \{ v , C_1 v ,C_2 v ,  A_1 C_2 v \}_{1,2,1} , \{ v , C_1 v , C_2 v , A_2 C_1 v , A_1 C_2 v \}_{1,2,2} \, .
\end{gather}
Note that the only difference with the case of shifted superpotential is for modules with dimension vector $\mbf d = (1,1,1)$.

BPS states of the quiver quantum mechanics are associated to irreducible quiver representations. However in the presence of an automorphism $\sigma$ we are supposed to keep only those states which are invariant under $\sigma$ \cite{Cecotti:2012va}. Such states correspond to \textit{invariant irreducible} representations, or $ii$-representations. The technical problem we will have to face has its origin in the fact that when computing BPS degeneracies using virtual localization techniques we do not deal directly with representations but with $\torus$-fixed points. This is because ``counting'' representations in the appropriate sense involves integrating over their moduli spaces.

To be more concrete, the $\zed_2$ automorphism $\sigma$ of the quiver  $\sfQ$ induces a functor $\mathscr{F}_\sigma$ from the category of quiver representations $\mathsf{rep} (\sfQ , \sfW)$ to itself. 
By using the correspondence between representations of $(\sfQ , \sfW)$ and left $\mathscr{J}_\cW$ modules, we can study the action of $\mathscr{F}_\sigma$ on the category $\mathscr{J}_\cW-\textsf{mod}$. Consider now a module $M$. We will denote its transform under the functor $\scrF_\sigma$ as $^{\sigma}M$. An isomorphic invariant indecomposable ($ii$-indecomposable) module, is then a module $M$ such that $M \simeq {^\sigma}M$. In our case, this can happen in precisely two cases: or $M$ is already invariant under $\scrF_\sigma$, or it can be used to construct the $ii$-module $N \simeq M \oplus {^\sigma}M$. Note that such $N$ is not indecomposable as an ordinary module, but only as an $ii$-module. This construction is familiar from quantum mechanics and it amounts in constructing a symmetric wave function, which describes a bosonic state\footnote{There is in principle the possibility of the corresponding particle state being described by a different statistics; it does not appear to be the case in all the examples we have studied but we don't have a clear argument for this.}. 

Here however we run into a problem with our formalism. To construct an $ii$-module of the form $M \oplus {^\sigma}M$ we have to symmetrize $M$ respect to the action of $\scrF_\sigma$. This is a different operation than symmetrizing each toric fixed point separately. For example when computing Euler characteristics by counting torus fixed points, one cannot tell the difference between two points and a $\PP^1$, which has two torus fixed points, the north and the south poles. The interplay between localization and $ii$-modules is a bit subtle, and require some additional information about the moduli spaces. We will now proceed to show how this problem can be solved, at least in our case.

In our example the cyclic modules $\{ v  \}_{1,0,0} , \{ v , C_1 v  , C_2 v \}_{1,2,0} , \{ v , C_1 v , C_2 v , A_2 C_1 v , A_1 C_2 v \}_{1,2,2} $ correspond to invariant representations: each cyclic module is clearly invariant under $\sigma$ and the virtual tangent space is trivial. The moduli space is just a point and each module correspond to a hypermultiplet.

The two modules $ \{ v , C_1 v \}_{1,1,0} , \{ v , C_2 v \}_{1,1,0}$ are exchanged by $\sigma$. However these cyclic modules correspond to fixed points in the moduli space of a $\PP^1$ family of representations. The virtual localization formula effectively averages over all the configurations and the two modules correspond to a representation which is already invariant under $\sigma$. This can be seen easily by looking at the Kronecker subquiver with arrows $C_1$ and $C_2$. The $\PP^1$ in question is precisely the representation spanned by maps of the form $[C_1 : C_2]$ (by partial abuse of language we use the same letters as in the cyclic modules). While each representation is not separately invariant, the operation of integrating over the moduli space produces an invariant and therefore corresponds to a physical BPS state. To see this more rigorously one can just compute the K-theoretic invariant; the computation is exactly the same as in the previous subsection and the result $\frac{1}{y} (1 + y^2)$ precisely correspond to the motivic class $[\PP^1] = \mathrm{pt} + [\motive]$ up to the overall normalization.

The same arguments can be repeated verbatim for the cyclic modules $ \{v , C_1 v , A_2 C_1 v , C_2 v \} _{1 ,2 ,1}$ and $\{ v , C_1 v ,C_2 v ,  A_1 C_2 v \}_{1,2,1}$, which together correspond to a $\PP^1$ family of representations.

The situation is however different for the cyclic modules $\{ v , C_2 \, v , A_1 \, C_2 v \}_{1,1,1} ,  \{ v , C_1 \, v , A_2 \, C_1 v \}_{1,1,1}$. These two modules are exchanged by the automorphism $\sigma$. By looking at the virtual tangent space (the computations are identical to those of the previous subsection) one sees immediately that the respective moduli spaces are points, corresponding to the cyclic modules themselves. Therefore the $ii$-representation is now described by the symmetrized module
\be
\{ v , C_2 \, v , A_1 \, C_2 v \}_{1,1,1} \bigoplus \{ v , C_1 \, v , A_2 \, C_1 v \}_{1,1,1} \, .
\ee
Note that this case is fundamentally different from the previous cases where the two cyclic modules correspond to fixed points of the \textit{same} (connected component of the) moduli space, in that cases a $\PP^1$. 

This $ii$-representation correspond physically to a \textit{single} stable BPS state, obtained by symmetrizing a wave function. Since it has no moduli this state must be an hypermultiplet. However a naive counting of the degeneracies via direct localization appears to give a degeneracy of 2. The solution to this puzzle is that to properly count the number of states associated to this representation we have to deal properly with the automorphism $\sigma$. A natural way of doing so is to consider  \textit{orbifold} Euler characteristics. Indeed the above module has precisely the form of a sum over images of the $\zed_2$ action. According to the definition of an orbifold Euler characteristic, to avoid overcounting we should divide by the order of the orbifold group, in this case $| \zed_2| = 2$. 

Therefore this computation reproduces the correct degeneracies with the superpotential \eqref{SU2-3-symmW}. Note however that this was possible since the involved geometries were simple enough. We could understand how to construct $ii$-modules only by a careful understanding of the moduli spaces, and the choice of physical combinations was done by hand. This procedure would become cumbersome in more complicated situations. At present we do not know how to implement sistematically the localization computation on $ii$-modules. A practical solution, as we have already seen and as used in \cite{Cirafici:2017wlw}, is to shift some of the variables in the superpotential as so to break as many automorphisms as possible. In that case one does not have to worry about $ii$-representation and the localization computation proceeds as usual.

\subsection{A dyonic line defect}

As our next example we will consider a dyonic line operator in a pure $\rm SU(3)$ gauge theory. We consider a region of the moduli space $\cal{B}$ where the relevant \textsc{BPS} quiver is given by
\begin{equation}
\xymatrix@C=8mm{
&  \bullet_1   \ar@{..>}@/^1.6pc/[rrrr] \ar@{..>}@/^1.8pc/[rrrr]  \ar@<-0.5ex>[rr]_{d_2}  \ar@<0.5ex>[rr]^{c_2}& &   \circ_2   \ar@<-0.5ex>[lldd]_{r}  \ar@<0.5ex>[lldd]^{s}  & & \ar@{..>}[ll] f \\
 & & & \\
& \circ_1  \ar[uu]^{b_1}  \ar@<-0.5ex>[rr]_{d_1}  \ar@<0.5ex>[rr]^{c_1} & &  \bullet_2  \ar[uu]_{a_1}   \ar@{..>}[rruu]  & & 
}
\end{equation}
with superpotential $\cW = r \left( a_1  c_1 - c_2  b_1 \right) + s \left( a_1  d_1 - d_2 b_1 \right)$.

The theory is coupled to a line defect with charge $\gamma_f = \gamma_{\bullet_2} + \gamma_{\circ_2}$. The new coupling determines a new superpotential term $\cW_L$, which can be naturally taken to be the same as above where the arrows $a_1$, $c_2$ and $d_2$ are replaced by the terms we can construct by composing the framing arrows. However the new terms can effectively be ignored, either because of the F-term relations from $\cW$ or because of the $\dim_\complex V_f = 1$ condition\footnote{This condition effectively tells us that we are only interested in representations for which the arrows going to the framing node are represented trivially. Therefore these arrows can be set to zero in the F-term relations and don't contribute to the construction of the moduli spaces.}. In this case the computation of the framed BPS state spectrum $\underline{\overline{\Omega}} (\gamma)$, including the toric action and the classification of the fixed points, essentially reduces to the one discussed in \cite{Chuang:2013wt}. We will therefore just take their results as a starting point and show how our formalism allows for the computation of the \textit{refined} spectrum. 

We define a toric action as $X \longrightarrow t_X \, X$ for each morphism $X$, which represents a field in the quiver quantum mechanics. To make this action compatible with the equations of motion we impose the conditions
\begin{align} \label{SU3weights1}
t_{a_1} t_{c_1} = t_{c_2} t_{b_1} \, , \qquad t_{a_1} t_{d_1} = t_{d_2} t_{b_1} \, , \qquad t_{c_1} t_{r_1} = t_{d_1} t_{s_1} \, , \qquad t_{r_1} t_{c_2} = t_{s_1} t_{d_2} \, ,
\end{align}
while the superpotential carries weight 
\be \label{SU3weights2}
t_{r_1} t_{a_1} t_{c_1} = t_{r_1} t_{c_2} t_{b_1} = t_{s_1} t_{a_1} t_{d_1} = t_{s_1} t_{d_2} t_{b_1} = \kappa \ .
\ee

$\torus$-fixed points were classified in \cite{Chuang:2013wt} and will be discussed momentarily. Around each fixed point $\pi$, the local structure of the moduli space is captured by the deformation complex constructed in \cite{Chuang:2013wt}
\begin{equation}
\xymatrix@C=8mm{  0 \ar[r] & \mathsf{S}^0_{\pi} \ar[r]^{\delta_0} & \mathsf{S}^1_\pi \ar[r]^{\delta_1} & \mathsf{S}^2_\pi \ar[r]^{\delta_2} & \mathsf{S}^3_\pi \ar[r] & 0
} \ .
\end{equation}
The relevant terms\footnote{For simplicity we omit those terms associated with the framing arrows which can be neglected in the localization computation.} are
\begin{align}
\mathsf{S}^0_\pi &=\mathrm{Hom}_{\mathbb{C}} (V_{\circ_1 , \pi} , V_{\circ_1 , \pi}) \oplus \mathrm{Hom}_{\mathbb{C}} (V_{\bullet_1 , \pi} , V_{\bullet_1 , \pi}) \oplus
\mathrm{Hom}_{\mathbb{C}} (V_{\circ_2 , \pi} , V_{\circ_2 , \pi}) \oplus \mathrm{Hom}_{\mathbb{C}} (V_{\bullet_2 , \pi} , V_{\bullet_2 , \pi}) \, ,
\cr
\mathsf{S}^1_\pi &= \mathrm{Hom}_{\mathbb{C}} (V_{\bullet_1 , \pi} , V_{\circ_2 , \pi}) \otimes (t_{c_2} + t_{d_2}) \oplus \mathrm{Hom}_{\mathbb{C}} (V_{\circ_1 , \pi} , V_{\bullet_2 , \pi}) \otimes (t_{c_1} + t_{d_1})  
\cr &
\oplus \mathrm{Hom}_{\mathbb{C}} (V_{\circ_1 , \pi} , V_{\bullet_1 , \pi}) \otimes (t_{b_1}) 
\oplus \mathrm{Hom}_{\mathbb{C}} (V_{\bullet_2 , \pi} , V_{\circ_2 , \pi}) \otimes (t_{a_1})
\cr &
\oplus \mathrm{Hom}_{\mathbb{C}} (V_{\circ_2 , \pi} , V_{\circ_1 , \pi}) \otimes (t_{r_1} + t_{s_1})
\oplus \mathrm{Hom}_{\mathbb{C}} (V_{f , \pi} , V_{\circ_2 , \pi}) \otimes (t_{f})    
\end{align}
The virtual tangent space has again the structure
\begin{equation} \label{TvirSU2}
\mathsf{T}^{\rm vir}_\pi = - \mathsf{S}^0_\pi + \mathsf{S}^1_\pi - \mathsf{S}^2_\pi + \mathsf{S}^3_\pi 
= - \mathsf{S}^0_\pi + \mathsf{S}^1_\pi - \mathbf{\kappa} \left(- \overline{\mathsf{S}^0_\pi} + \overline{\mathsf{S}^1_\pi} \right) = \sum_i w_i - \sum_i \frac{\kappa}{w_i}
\end{equation}
where we have chosen the weight of the framing arrow $f \dashrightarrow \circ_2$ equal to the weight of the superpotential, borrowing the results of \cite{Cirafici:2011cd}. This can be accomplished by choosing appropriately the weights of the other framing arrows.

The table \ref{tableFP} contains all the $\torus$-fixed cyclic modules, together with their dimension vectors $\mbf d = (1, d_{\circ_1} , d_{\circ_2} , d_{\bullet_1} , d_{\bullet_2})$ and the sum of weights in the virtual tangent space.
\begin{table}[h]
\def\arraystretch{1.4}
\begin{tabular}{|l|l|l|}
\hline
Dimension  & $\torus$-fixed module  & $\sum_i w_i$  \\ \hline
 $(1,0,0,0,0)$ & $\{ \emptyset \}$  & $0$  \\ \hline
$(1,0,1,0,0)$  &$ \{ v \}$  & $0$  \\ \hline
$(1,1,1,0,0)$  &$ \{ r \, v , v \}$  & $\frac{t_s}{t_r}$  \\ \hline
  &$ \{ s \, v ,  v \}$  & $\frac{t_r}{t_s}$  \\ \hline
$(1,2,1,0,0)$  &$ \{ s \, v , r \, v , v \}$  & $0$  \\ \hline
$(1,1,1,0,1)$  &$ \{ d_1 \, r \, v , r \, v , v \}$  & $\frac{t_{c_1}}{t_{d_1}} + t_a t_{d_1} t_r + \frac{t_s}{t_r}$  \\ \hline
  &$ \{ c_1 \, s \, v , s \, v ,  v \}$  & $\frac{t_{d_1}}{t_{c_1}} + \frac{t_r}{t_s} + t_{a} t_{c_1} t_s$  \\ \hline
$(1,2,1,0,1)$  &$ \{ d_1 \, r \, v , s \, v , r \, v , v \}$  & $\frac{t_{c_1}}{t_{d_1}} + t_a t_{d_1} t_r + \frac{t_s}{t_r} + \frac{t_{c_1} t_s}{t_{d_1} t_r}$  \\ \hline
  &$\{ c_1 \, r \, v , s \, v , r \, v , v \}$  & $\frac{t_{d_1}}{t_{c_1}} + \frac{t_s}{t_r}$  \\ \hline
  &$ \{ c_1 \, s \, v , s \, v , r \, v , v \}$  & $\frac{t_{d_1}}{t_{c_1}} + \frac{t_r}{t_s} + \frac{t_{d_1} t_r}{t_{c_1} t_s}+ t_a t_{c_1} t_s$  \\ \hline
$(1,2,1,0,2)$  &$ \{c_1 \, r \, v ,  d_1 \, r \, v , s \, v , r \, v , v \}$  & $t_a t_{d_1} t_r + 2 \frac{t_s}{t_r} + \frac{t_{c_1} t_s}{t_{d_1} t_r}$  \\ \hline
  &$\{ c_1 \, s \, v , d_1 \, r \, v , s \, v , r \, v , v \}$  & $\frac{t_{c_1}}{t_{d_1}} + \frac{t_{d_1}}{t_{c_1}}+ t_a t_{d_1} t_r + t_a t_{c_1} t_s + \frac{t_r}{t_s}+ \frac{t_s}{t_r}$  \\ \hline
  &$ \{c_1 \, s \, v , c_1 \, r \, v , s \, v , r \, v , v \}$  & $2 \frac{t_{d_1}}{t_{c_1}} + \frac{t_{d_1} t_r}{t_{c_1} t_s} + t_a t_{c_1} t_s$  \\ \hline
$(1,2,1,0,3)$  &$ \{c_1 \, s \, v , c_1 \, r \, v ,  d_1 \, r \, v , s \, v , r \, v , v \}$  & $\frac{t_{d_1}}{t_{c_1}} + t_a t_{d_1} t_r + t_a t_{c_1} t_s + \frac{t_s}{t_r}$  \\ \hline
\end{tabular}
\caption{The table contains the list of the $\torus$-fixed modules, their dimensions and the information about their toric weights}
\label{tableFP}
\end{table}
Every $\torus$-fixed point whose virtual tangent space has the structure $\mathsf{T}_{\pi}^{\rm vir}= \sum_i w_i - \sum_i \kappa / w_i$ will contribute a term $\hat{\sf a} \left(  \sum_i w_i - \sum_i \kappa / w_i \right) $ to the BPS generating function. Let us examine the nontrivial contributions from the above table one by one. 

\begin{enumerate}
\item 
\underline{$\mathbf{d} = (1,1,1,0,0)$}. From the table we simply have the sum
\begin{align}
\hat{\sf a} \left(  \frac{t_s}{t_r} - \kappa \frac{t_r}{t_s} \right) + \hat{\sf a} \left(  \frac{t_r}{t_s} - \kappa \frac{t_s}{t_r} \right) = \frac{t_s-\kappa t_r}{\sqrt{\kappa} (t_r-t_s)} + \frac{\kappa t_s-t_r}{\sqrt{\kappa} (t_r-t_s)} = -\left( \sqrt{\kappa} + \frac{1}{\sqrt{\kappa}} \right)
\end{align}
\item 
\underline{$\mathbf{d} = (1,1,1,0,1)$}. Again we have two fixed points. To simplify the computation we use the relations \eqref{SU3weights1} and \eqref{SU3weights2} between the toric weights to write 
\begin{align}
\frac{t_{c_1}}{t_{d_1}} + t_a t_{d_1} t_r + \frac{t_s}{t_r} &=  2 \frac{t_s}{t_r} + \kappa \frac{t_r}{t_s} \cr
\frac{t_{d_1}}{t_{c_1}} + \frac{t_r}{t_s} + t_{a} t_{c_1} t_s &=  2 \frac{t_r}{t_s} + \kappa \frac{t_s}{t_r} 
\end{align}
Therefore
\begin{align}
\hat{\sf a} \left(  2 \frac{t_s}{t_r} + \kappa \frac{t_r}{t_s} - \kappa \left( 2 \frac{t_r}{t_s} + \frac{1}{\kappa} \frac{t_s}{t_r}  \right) \right) + \hat{\sf a} \left( 2 \frac{t_r}{t_s} + \kappa \frac{t_s}{t_r} - \kappa \left( 2 \frac{t_s}{t_r} + \frac{1}{\kappa} \frac{t_r}{t_s} \right) \right) = - \left( \sqrt{\kappa} + \frac{1}{\sqrt{\kappa}} \right)
\end{align}
\item 
\underline{$\mathbf{d} = (1,2,1,0,1)$}. Proceeding as before we have
\begin{align}
\frac{t_{c_1}}{t_{d_1}} + t_a t_{d_1} t_r + \frac{t_s}{t_r} + \frac{t_{c_1} t_s}{t_{d_1} t_r} &= 2 \frac{t_s}{t_r} + \kappa \frac{t_r}{t_s} + \frac{t_s^2}{t_r^2} \cr
\frac{t_{d_1}}{t_{c_1}} + \frac{t_s}{t_r} &= \frac{t_s}{t_r} + \frac{t_r}{t_s} \cr
\frac{t_{d_1}}{t_{c_1}} + \frac{t_r}{t_s} + \frac{t_{d_1} t_r}{t_{c_1} t_s} &= 2 \frac{t_r}{t_s} + \frac{t^2_r}{t_s^2} + \kappa \frac{t_s}{t_r}
\end{align}
Therefore, denoting by $\pi_i$ the three fixed points, we easily see that
\be
\sum_{i=1}^3 \hat{\sf a} \left( T_{\pi_i}^{\rm vir} \right) = \frac{1}{\kappa} + 1 + \kappa
\ee
\item 
\underline{$\mathbf{d} = (1,2,1,0,2)$}. Again we can use the relations between the toric weights to simplify
\begin{align}
t_a t_{d_1} t_r + 2 \frac{t_s}{t_r} + \frac{t_{c_1} t_s}{t_{d_1} t_r} &= \kappa \frac{t_r}{t_s} + 2 \frac{t_s}{t_r} + \frac{t_s^2}{t_r^2} \cr 
\frac{t_{c_1}}{t_{d_1}} + \frac{t_{d_1}}{t_{c_1}}+ t_a t_{d_1} t_r + t_a t_{c_1} t_s + \frac{t_r}{t_s}+ \frac{t_s}{t_r} &=
2 \frac{t_s}{t_r} + 2 \frac{t_r}{t_s} + \kappa \frac{t_r}{t_s}+ \kappa \frac{t_s}{t_r}  \cr
 2 \frac{t_{d_1}}{t_{c_1}} + \frac{t_{d_1} t_r}{t_{c_1} t_s} + t_a t_{c_1} t_s &= 2 \frac{t_r}{t_s} + \frac{t_r^2}{t_s^2} + \kappa \frac{t_s}{t_r}
\end{align}
If we denote by $\pi_i$ the three fixed points, we have again
\be
\sum_{i=1}^3 \hat{\sf a} \left( T_{\pi_i}^{\rm vir} \right) = \frac{1}{\kappa} + 1 + \kappa
\ee
\item 
\underline{$\mathbf{d} = (1,3,1,0,2)$}. Finally we have
\be
\frac{t_{d_1}}{t_{c_1}} + t_a t_{d_1} t_r + t_a t_{c_1} t_s + \frac{t_s}{t_r} = \frac{t_r}{t_s} + \frac{t_s}{t_r} + \kappa \frac{t_r}{t_s} + \kappa \frac{t_s}{t_r}
\ee
We see easily that the weights are paired and the contribution of this fixed point is simply $1$.
\end{enumerate}
To conclude we introduce the refined variable $y = - \sqrt{\kappa}$. For the quantum line operator we have found the prediction
\begin{align}
\mathscr{L}_{\gamma_f} & = \sum_{d = (1 , d_\circ )} \chi \left( \mathscr{M}_\gamma  , \cO^{\rm vir} \otimes \cK^{1/2}_{\rm vir} \right) \sfX_{\gamma_f + d_{\circ_1} \gamma_{\circ_1}+d_{\circ_2} \gamma_{\circ_2}+ d_{\bullet_1} \gamma_{\bullet_1} + d_{\bullet_2} \gamma_{\bullet_2}} \\
& = \sfX_{\gamma_f} +
\sfX_{\gamma_f + \gamma_{\circ_2}} +
\left( y + \frac{1}{y} \right) \sfX_{\gamma_f + \gamma_{\circ_1}+ \gamma_{\circ_2}} +
\sfX_{\gamma_f + 2 \gamma_{\circ_1}+ \gamma_{\circ_2}} + 
\left( y + \frac{1}{y} \right) \sfX_{\gamma_f + \gamma_{\circ_1}+ \gamma_{\circ_2}+ \gamma_{\bullet_2}} 
\cr \nonumber & \ \ + 
\left( \frac{1}{y^2} + 1 + y^2 \right) \sfX_{\gamma_f +2  \gamma_{\circ_1}+ \gamma_{\circ_2}+ \gamma_{\bullet_2}}+ \left( \frac{1}{y^2} + 1 + y^2 \right)
\sfX_{\gamma_f +2 \gamma_{\circ_1}+ \gamma_{\circ_2}+ 2 \gamma_{\bullet_2}} +
\sfX_{\gamma_f +2  \gamma_{\circ_1}+ \gamma_{\circ_2}+ 3 \gamma_{\bullet_2}}
\end{align}
Note that at each step in the localization computation all the toric weights cancel, with the exception of $\kappa$. Also in this case there is no need to take a scaling limit to compute the refined BPS invariants.

\section{Discussion} \label{discussion}

In this note we have discussed a formalism which, upon certain data being known, takes a line operator in theories of class $\cS$ and computes the protected spin characters of its framed BPS states.  This is obtained by associating to the line operator the Euler characteristic of a complex of sheaves, which is then evaluated by localization. The protected spin characters are then identified with refined Donaldson-Thomas invariants. 

The main idea is to adapt the formalism constructed by Nekrasov and Okounkov \cite{Nekrasov:2014nea} in their study of M2 branes by means of the K-theoretic version of Donaldson-Thomas theory. This formalism has a natural interpretation as a step towards a categorification of BPS spectra, in that it replaces the numerical Donaldson-Thomas invariant with the Euler characteristic of a complex of sheaves. Such complex appears as a resolution of a certain structure sheaf associated with the moduli space. This can be made very concrete working equivariantly with respect to a natural toric action, where all the ingredients appear naturally in the virtual tangent space.

In this note we have modified this construction to deal with a particular class of quivers, which describe the IR coupling of a line defect to a quantum field theory of class $\cS$. To ensure that the enumerative problem is well defined we have limited ourselves to quivers which have already been studied in \cite{Chuang:2013wt,Cirafici:2018jor,Cirafici:2017wlw}. For these quivers (and those which can be obtained by decoupling limits) the formalism can be applied to determine the spin content of the framed BPS spectra and therefore the quantum line operators. The protected spin characters can then be interpreted as a sum of certain modules, generated by the quantum parameter $y$. Furthermore we have also described how to set up the localization computation in the presence of non-trivial quiver automorphisms, clarifying some assumptions which where implicit in \cite{Cirafici:2017wlw}.

The same formalism can be adapted to Calabi-Yau quivers which arise as singular limit of topological string compactifications and will appear in a separate paper \cite{MQ}.

Our construction is purely algebraic. What is certainly lacking is a geometrical interpretation of the relevant sheaves and quantum parameters. It should be possible to obtain a more concrete geometrical picture using the full formalism of  \cite{Chuang:2013wt}, which consists in engineering the problem via D-branes wrapping cycles in a Calabi-Yau and then taking an appropriate IR limit to decouple string states. For example this should relate the quantum parameter $y$ to the square root of the canonical bundle over the Calabi-Yau, as in \cite{Nekrasov:2014nea}. Furthermore it would be interesting to relate our construction with the more geometrical one of \cite{Moore:2015szp,Brennan:2018yuj}. 

In particular if the line operator is associated with a loop of the form ${\rm pt} \times \wp$ on $\real \times \cC$ together with an irreducible representation, our results associate to such a loop a K-theoretic enumerative problem  and certain complexes of sheaves. It would be very interesting to pursue this avenue further in order to understand if our construction could be related to a version of Khovanov homology for such loops on $\real \times \cC$. In particular our formalism arises very naturally from the perspective of the $\cN=(2,0)$ superconformal theory and it would be interesting to have a more in depth comparison with \cite{Witten:2011zz}.

A more ambitious question would be to use our formalism to investigate the algebra of line operators. From our perspective the K-theoretic computation should directly replace the algebra coefficients with equivariant modules. It would be very interesting to work out the details and understand the relation with the cohomological Hall algebra of \cite{Kontsevich:2010px}. In this respect an intermediate step would be to understand the relation between our construction and \cite{Galakhov:2018lta}.
\section*{Acknowledgements}

I am grateful to Michele del Zotto, Vivek Shende, Yan Soibelman, Vasily Pestun and Johannes W\"alcher for discussions. I am thankful to the organizers of the program \textit{Symplectic Geometry and Representation Theory} at the Hausdorff Institute for Mathematics in Bonn for the warm hospitality during the last stages of this project. These results were presented at the workshops \textit{Geometry and Topology inspired by Physics} in 2018 in Ascona and \textit{Young Researchers in String Mathematics} in 2017 in Bonn, and I am grateful to the organizers for the invitation to speak and for the warm hospitality. I am a member of INDAM-GNFM, I am supported by INFN via the Iniziativa Specifica GAST and by the FRA2018 project ``K-theoretic Enumerative Geometry in Mathematical Physics''.

\end{document}